\documentclass[aps,pre,reprint,superscriptaddress,nofootinbib]{revtex4-1}
\usepackage{graphicx}
\usepackage{float}
\usepackage{csquotes}

\usepackage{mathtools} 

\usepackage[usenames,dvipsnames]{color}
\usepackage[normalem]{ulem}

\begin{document}


\title{Adaptive stochastic continuation with a modified lifting procedure applied to complex systems}


\author{Clemens Willers}
\email[]{clemens.willers@uni-muenster.de}
\thanks{ORCID ID: 0000-0001-5777-0514}
\affiliation{Institut f\"ur Theoretische Physik, Westf\"alische Wilhelms-Universit\"at M\"unster, 48149 M\"unster, Germany}
\affiliation{Center for Nonlinear Science (CeNoS), Westf{\"a}lische Wilhelms-Universit\"at M\"unster, 48149 M\"unster, Germany}
\author{Uwe Thiele}
\email{u.thiele@uni-muenster.de}
\homepage{http://www.uwethiele.de}
\thanks{ORCID ID: 0000-0001-7989-9271}
\affiliation{Institut f\"ur Theoretische Physik, Westf\"alische Wilhelms-Universit\"at M\"unster, 48149 M\"unster, Germany}
\affiliation{Center for Nonlinear Science (CeNoS), Westf{\"a}lische Wilhelms-Universit\"at M\"unster, 48149 M\"unster, Germany}
\affiliation{Center for Multiscale Theory and Computation (CMTC), Westf{\"a}lische Wilhelms-Universit\"at, 48149 M\"unster, Germany}
\author{Andrew J. Archer}
\thanks{ORCID ID: 0000-0002-4706-2204}
\affiliation{Department of Mathematical Sciences, Loughborough University, Loughborough LE11 3TU, United Kingdom}
\affiliation{Interdisciplinary Centre for Mathematical Modelling, Loughborough University, Loughborough LE11 3TU, United Kingdom}
\author{David J. B. Lloyd}
\thanks{ORCID ID: 0000-0002-1902-0007}
\affiliation{Department of Mathematics, University of Surrey, Guildford, GU2 7XH, United Kingdom}
\author{Oliver Kamps}
\thanks{ORCID ID: 0000-0003-0986-0878}
\affiliation{Center for Nonlinear Science (CeNoS), Westf{\"a}lische Wilhelms-Universit\"at M\"unster, 48149 M\"unster, Germany}


\date{\today}

\begin{abstract}
Many complex systems occurring in the natural or social sciences or economics are frequently described on a microscopic level, e.g., by lattice- or agent-based models. To analyze the states of such systems and their bifurcation structure on the level of macroscopic observables, one has to rely on equation-free methods like stochastic continuation. Here, we investigate how to improve stochastic continuation techniques by adaptively choosing the parameters of the algorithm. This allows one to obtain bifurcation diagrams quite accurately, especially near bifurcation points. We introduce lifting techniques which generate microscopic states with a naturally grown structure, which can be crucial for a reliable evaluation of macroscopic quantities. We show how to calculate fixed points of fluctuating functions by employing suitable linear fits. This procedure offers a simple measure of the statistical error. We demonstrate these improvements by applying the approach in analyses of (i) the Ising model in two dimensions, (ii) an active Ising model, and (iii) a stochastic Swift-Hohenberg model. We conclude by discussing the abilities and remaining problems of the technique.

\end{abstract}


\maketitle

\section{Introduction\label{introduction}}

In many fields of science -- such as physics, chemistry, biology, economics, and sociology --  the behavior of various complex systems is captured by nonlinear mathematical models that describe the time evolution of the system. The models are formulated in terms of state variables and depend on specific parameters. They can exhibit various stable and unstable solutions corresponding, e.g., to steady states, time-periodic states or chaotic states of the system. Such a general classification holds for model types ranging from discrete agent-based models to continuum models. It directly applies to deterministic models while for stochastic models it applies when considering certain statistical quantities \cite{DeutschMainiDormann2007, ChopardDroz2005b, Tu1994, Haken2006, Pismen2006, GlansdorffPrigogine}. In this context, we regard molecular dynamics (MD) simulations and similar systems as stochastic systems. Although formally they are deterministic, in practice their very many degrees of freedom cause deterministic chaos that renders them stochastic \cite{PhysRevLett.65.1391, PhysRevLett.69.3306, PhysRevLett.72.1644}.

Nonlinear models describe a rich variety of qualitatively different system behavior and related (possibly abrupt) transitions that may occur when conditions change:  States change their stability, additional states emerge or several states exist at identical parameter values (multistability, hysteresis). For instance, much attention in the fields of natural science and engineering has focused on ``tipping points'' where small changes of a control parameter result in 
large changes in system behavior \cite{TippingPointsAWVC2012}.
All this information can be summarized in \textit{bifurcation diagrams} that present how the various states of the studied system or solutions of the developed model emerge, change, and vanish when varying a control parameter \cite{Kuznetsov2010,Strogatz2014,ArgyrisFaustHaaseFriedrich2015}. For this, a real quantity is defined as the order parameter or solution measure which characterizes and distinguishes the different states. This quantity -- which can, e.g., be a physical observable, a statistical quantity, or a norm -- is plotted against the control parameters \cite{poincare1885,LandauPhaseTransitions}. Here we restrict ourselves to analyzing steady states of the system, i.e., the solution measure does not involve averaging over time.

For deterministic models consisting, e.g., of systems of ordinary or partial differential equations, a well-developed and efficient tool for calculating bifurcation diagrams is \textit{numerical path continuation} \cite{KrauskopfOsingaGalan-Vioque2007, DWCD2014ccp, AllgowerGeorg1987}. With these methods one can directly follow branches of solutions and detect changes in their stability and, in consequence, bifurcations where additional solution branches emerge that one can switch onto. In contrast to procedures based on numerical time simulations, continuation can detect and follow branches of both stable and unstable states. It is used in many examples from classical dynamical systems \cite{Kuznetsov2010, ArgyrisFaustHaaseFriedrich2015} to spatially extended systems such as those occurring in fluid dynamics or reaction-diffusion systems \cite{Mei2000, SaML2002jcp, DWCD2014ccp, SaNe2016epjt, EGUW2019springer}.

To apply numerical continuation, one normally needs an evolution equation at the particular scale of interest in a closed form. That is, for the order parameter $X$ and a parameter $\lambda$ (we only regard one-dimensional parameter spaces), one considers a differential equation,
\begin{equation}\label{eq:dglI}
 \dot{X} = f_{\lambda}(X)
\end{equation}
or an iterative mapping,
\begin{equation}\label{eq:imI}
 X_{t+\tau} = F_{\lambda}(X_t)
\end{equation}
where $\tau$ is the time step of the mapping.
Here we are interested in steady states $X_s(\lambda)$ at different values of $\lambda$, i.e., in solutions of the equation $f_{\lambda}(X_s(\lambda)) = 0$ or $F_{\lambda}(X_s(\lambda)) = X_s(\lambda)$, respectively. Note that a differential equation can be transformed into an iterative map using time discretization and integration. So, generally speaking, we need to obtain the roots of a function
\begin{equation}\label{eq:G}
 G_{\lambda}(X_s(\lambda)) := F_{\lambda}(X_s(\lambda)) - X_s(\lambda) = 0.
\end{equation}
Hence, continuation means finding zeros of the function $G_{\lambda}$ for subsequent values of the parameter $\lambda$. A complete bifurcation diagram is calculated by following all solution branches step by step. In every step, zeros obtained in the previous step or several previous steps can be used to predict a next value. This value is then used as initial guess for the correcting root finding procedure (see, e.g., chapter 10 of Ref.~\cite{Kuznetsov2010}). Advancing step by step in $\lambda$ in this prediction-correction modus represents the simplest realization of numerical path continuation.

However, there exist many examples where such equations are only implicitly given. In this case, one can still apply path continuation in an \textit{equation-free} way. That is, one uses a \textit{time stepper} instead of explicit evolution equations \cite{ThQK2000pnasusa,Kev2009,ConsumerLockIn,Kueh2012sjsc,ThLS2016pa,barkley2006}. This technique is called \textit{stochastic continuation} and makes it possible to apply continuation techniques for a much wider range of systems than those captured by Eqs.~(\ref{eq:dglI}) or (\ref{eq:imI}), including real-world experiments~\cite{Sieber2007,Sieber2008,Barton2012}.

Typical examples where the equation-free approach is applied are lattice-based or agent-based models \cite{ConsumerLockIn,ThLS2016pa}, MD simulations \cite{FSVL2009DCDSB} and kinetic Monte Carlo methods \cite{MaMK2002jcp}. In all these examples, the model is defined on a microscopic level whereas one is interested in the dynamics of a macroscopic quantity for which an analytical description is not available. The role of the macroscopic quantity is often taken by a statistical quantity, i.e., one analyzes ensembles of states. This is the origin of the term \textit{stochastic continuation}, i.e., the method is not limited to stochastic contexts. Variants are also applied to stochastic partial differential equations  \cite{Kueh2015sjuq}. 

In summary, stochastic continuation refers to continuation methodologies that one can apply without having direct access to the dynamics of the quantity of interest. An indirect access is made possible by three steps: \textit{lifting}, \textit{evolving}, and \textit{restricting}. First, in the lifting step one creates a suitable microscopic state (or an ensemble of such states) belonging to a specific value of the macroscopic quantity. Then, the state is evolved via a \textit{time stepper}, i.e., the microscopic time evolution is advanced. Finally, in the restricting step the macroscopic quantity belonging to the newly evolved microscopic state is calculated. Taken together, these three steps realize the required time evolution on the macroscopic level and replace the function $F_{\lambda}$.

The present work improves several aspects of the stochastic continuation method. To begin with, in most cases, the output of the time stepper in the evolving step contains statistical fluctuations. We propose an alternative way of handling this issue. Namely instead of directly applying a root finding method to the fluctuating function $G_{\lambda}$, we propose to evaluate a few (between 10 and 50) function values in a suitable neighborhood of the initial guess and perform a linear fit; see Ref.~\cite{Pasupathy2011}. The root of the linear fit is found to be a good approximation for the desired steady state. This procedure moderates the influence of fluctuations and works in a simple and stable manner.

Furthermore, the stochastic continuation algorithm involves quite a large number of numerical parameters, e.g., the length of the microscopic time evolution in the evolving step, the number of microscopic realizations employed, and the step size of the path continuation step. The corresponding parameter settings have a crucial influence on the quality and precision of the continuation result. In Ref.~\cite{ThLS2016pa}, one possible way to systematically determine these parameters is described. It is based on an analysis of the probability distribution of the order parameter and how it changes during the microscopic time evolution step. This analysis is done during a configuration step prior to the continuation run. As a further improvement, we suggest here to adaptively adjust certain specific parameters on the fly. Especially near bifurcation points, this leads to the results being more accurate.

We also consider the lifting procedure, which is a critical part of stochastic continuation. Being a one-to-many mapping, it is not uniquely defined. In most cases described in the literature, a simple constrained random lifting procedure is used, i.e., a microscopic state with the correct macroscopic observable is chosen randomly without any control of its internal structure \cite{ThLS2016pa, ConsumerLockIn, ThQK2000pnasusa, MaMK2002jcp}. Other lifting procedures found in the literature concern specific examples. In Ref.~\cite{FSVL2009DCDSB}, a lifting technique for MD simulations of dense fluids is described where the positions and velocities of atoms are chosen according to appropriate probability distributions based on the values of the macroscopic quantities. In addition, a detailed study of lifting errors for this particular situation is given. Another example involving MD simulations (water in carbon nanotubes) is discussed in Ref.~\cite{SrKH2005PRL}. There, the lifting is done using a constrained MD simulation with an artificial bias potential. In Ref.~\cite{KoPK2005JCP} a lifting for a grand canonical Monte Carlo simulation of the Larson model for micelle formation is proposed. The lifting is based on a database of microscopic structures which are obtained by performing lengthy time simulations.

Under the assumption that there is a clear separation of the slow macroscopic and the fast microscopic timescales, possible lifting errors can be \enquote{healed} during the microscopic time evolution performed during the evolving step. Considering an ensemble of microscopic states, a similar assumption is made: The ensemble is generated according to the low-order moment(s) of the appropriate distribution that are known from the macroscopic observable(s). The unknown higher-order moments are expected to establish their values much more rapidly than the low-order ones. Through this slaving, errors made in forming the ensemble are healed \cite{ThLS2016pa, MaMK2002jcp, Sieber2018}. The healing process is supposed to constitute a minor part of the algorithm's computation time.

However, in many important cases, the microscopic states show an internal spatial structure that is important for the macroscopic dynamics (see Sec.~\ref{sec:ising}). The healing process does develop these internal structures, but this can sometimes only occur after a very long microscopic time evolution, i.e., costing a huge numerical effort. In such cases, stochastic continuation has no advantage for stable states over calculations based on just direct time simulations.

In the literature, various ways to tackle this problem are presented. Reference~\cite{Kev2005} uses an algorithm which creates initial states for the evolving step that are very close to a \textit{slow manifold} of the system. However, this only works for specific coupled differential equations. In Ref.~\cite{Kev2003} a single suitable fixed reference state is used and adjusted according to the values of the macroscopic quantities. However, this does not allow the character of the microscopic states to change over the course of the continuation process and results in similar problems as a constrained random lifting. The lifting techniques used for the MD simulation of water in carbon nanotubes \cite{SrKH2005PRL} and of micelle formation \cite{KoPK2005JCP} mentioned above are able to produce naturally grown microscopic structures. Yet, they require a vast numerical effort because they are based on long-time simulations at every single lifting step and on the creation of a database of clustered structures, respectively.

Here, we propose a solution to this problem that consists of a lifting technique which uses a flexibly varying reference state. This facilitates the creation of suitably structured subsequent microscopic states for different values of the macroscopic quantity as one follows the continuation path. In particular, the continuation of stable branches gives much more accurate results with this procedure, which we refer to as \textit{structure lifting}.

After introducing our proposed improvements to the stochastic continuation procedure, we illustrate the approach by applying it to three distinct systems and discuss the results. This allows to focus on the strengths and weaknesses of our method, but we do not give an exhaustive analysis of the individual example systems.

The first system we apply the approach to is the Ising model in two dimensions (2D). This is a simple model for ferromagnetism \cite{baxter2016exactly}. It consists of a 2D lattice containing $N$ sites that are each occupied by a spin which can be in one of two possible states, $s_i\in\{-1,1\}$, with $i=1,\dots,N$. Every spin interacts with its four nearest neighbors with \textit{interaction strength} $J$. The Hamiltonian reads
\begin{equation}\label{eq:Ising_H}
	H(s) = -\sum_{\mathclap{\textrm{nn}}}Js_is_j
\end{equation}
where $s=\{s_1,s_2,\cdots,s_N\}$ and $\sum_\textrm{nn}$ denotes the sum over nearest-neighbor pairs. The macroscopic observable is the magnetization $m$ which is the mean value of any of the spins, i.e., $m:=\langle s_i \rangle$, where $\langle s_i \rangle$ denotes the time (or ensemble) average value of $s_i$. The microscopic dynamics of the model is given by the usual Metropolis Monte Carlo algorithm \cite{MRRT1953JCP}. One is interested in the stable and unstable states, which are characterized by their magnetization value, which is a function of the reduced temperature $T':=k_\mathrm{B}T/J$, where $T$ is the temperature and $k_\mathrm{B}$ is Boltzmann's constant. Onsager's analytical solution shows that above a critical temperature $T_c$ the magnetization is zero while the system shows a spontaneous magnetization below $T_c$ \cite{Onsa1944pr}.

The second example we consider is the active Ising model introduced in Ref.~\cite{SoTa2015pre}. Here the spins can be associated with particles that not only can flip their spin state but can also move. The preferred direction of motion of a spin depends on its orientation. Depending on the mean particle density, different phases are observed. At low densities, the mean orientation and the mean velocity are zero. This is called the \enquote{gas phase.} At high densities, the spins are ordered, i.e., there emerges an overall magnetization which is connected to an overall directed motion. This is called the \enquote{liquid phase.} At intermediate densities, a phase separation occurs, into different regions containing the gas and the liquid states. The total fraction containing the liquid can be used as the macroscopic order parameter (see Sec.~\ref{sec:activeIsing} for details).

Finally, as third example we perform stochastic continuation of the steady inhomogeneous solutions of a stochastic partial differential equation, namely a stochastic version of the Swift-Hohenberg equation \cite{Hutt2008el, HVTM1993prl}. It incorporates an additive noise field $\eta$ multiplied by a parameter $D$ which determines the strength of the noise term. The time evolution equation for the field $\Psi(x,y,t)$, which varies with position $(x,y)$ and time $t$, is
\begin{equation}\label{eq:sto_SH}
	\partial_t \Psi = \varepsilon\Psi - \Psi^3 - (1+\nabla^2)^2\Psi + \sqrt{D}\eta.
\end{equation}
Different implementations of stochasticity (i.e.,\ the noise field $\eta$) are employed in the literature: One finds additive Gaussian spatiotemporal noise \cite{VHST1991pra, ElVG1992prl, ACLR2013pre}, independent Gaussian additive and multiplicative noise \cite{TaNS2002pla}, and spatially global Gaussian noise \cite{Hutt2008el}. Here we take the noise to be normally distributed and uncorrelated in space and time. Depending on the control parameter $\varepsilon$ and the magnitude of $D$, this equation has various stable steady states which correspond, e.g., to stripe or square patterns of the field $\Psi$. A macroscopic quantity which can be used as a solution measure (order parameter) in a bifurcation diagram is the $L^2$ norm of the deviations of $\Psi$ from its mean value.

This paper is organized as follows: In Sec.~\ref{sec:method}, we introduce the details of stochastic continuation and the corresponding notations. Additionally, we explain some difficulties with the method and present our proposed improvements. We illustrate the improved method with three examples: the classical Ising model in 2D  in Sec.~\ref{sec:ising}, an active Ising model in Sec.~\ref{sec:activeIsing}, and a stochastic Swift-Hohenberg equation in Sec.~\ref{sec:SwiftHohenberg}. Finally,  in Sec.~\ref{sec:conclusion_outlook}, we summarize the capabilities of the proposed approach and discuss some remaining questions.

\section{Method\label{sec:method}}

\subsection{Numerical continuation and the pseudoarclength method}
\label{sec:numcont}

The method of stochastic continuation extends the standard equation-based continuation method for deterministic systems. Therefore, we start by briefly explaining the essence of continuation methods for one-dimensional deterministic systems \cite{KrauskopfOsingaGalan-Vioque2007,DWCD2014ccp,AllgowerGeorg1987}. These have a state variable (or order parameter) $X$ that is a real macroscopic quantity. The dynamics is defined by an ordinary differential equation (ODE) of the form in Eq.~(\ref{eq:dglI}) or by a iterative map of the form in Eq.~(\ref{eq:imI}), which depend on a control parameter $\lambda$ (we only consider one-dimensional parameter spaces). Note that a dynamics described by a partial differential equation (PDE) or an integrodifferential equation can be transformed into a system of ODEs, e.g., by discretizing in real space using finite differences or spectral methods. Further, one may employ time discretization and integration to transform a differential equation into an iterative map. Therefore, here we only consider the latter.

Of central interest is how steady states $X_s(\lambda)$ depend on the value of the control parameter $\lambda$; i.e.,\ we neglect time-periodic and chaotic states. In other words, the aim is to search for states which satisfy Eq.~(\ref{eq:G}). Presented in a bifurcation diagram as a function of $\lambda$, these zeros form different branches (this is a consequence of the implicit function theorem \cite{KrauskopfOsingaGalan-Vioque2007}). Applying the continuation method, each branch can be calculated by following it step by step. In every step, a predictor-corrector scheme is applied, whereby different prediction and correction methods are used in the literature.

In the simplest version of continuation, one explicitly changes the parameter $\lambda$ in small steps. The corresponding steady states $X_s(\lambda)$ are obtained by a root finding procedure applied to the function $G_{\lambda}$ in Eq.~(\ref{eq:G}) for a fixed value of $\lambda$. In this approach the solution at each step in $\lambda$ can be obtained by using the solution from the previous step as the initial guess for the solution at the next step. Thinking of this as a \enquote{prediction} followed by a \enquote{correction,} then the prediction step is trivial and is represented by a horizontal line in the bifurcation diagram, while the corrections correspond to vertical displacements.

Taking two previous steps into account and using a linear extrapolation, it is possible to use this secant predictor to obtain a better initial state in the predictor step. Again, the corrector is a root finding process at fixed $\lambda$. The correction at fixed $\lambda$ is problematic at saddle-node bifurcations because one cannot follow the bifurcation curve through the turning point (or fold). Instead, the predictor goes beyond the bifurcation point and the corrector either does not converge to any steady solution or converges to a solution on a branch different to the desired one.

\begin{figure}
  \includegraphics[width=0.9\hsize]{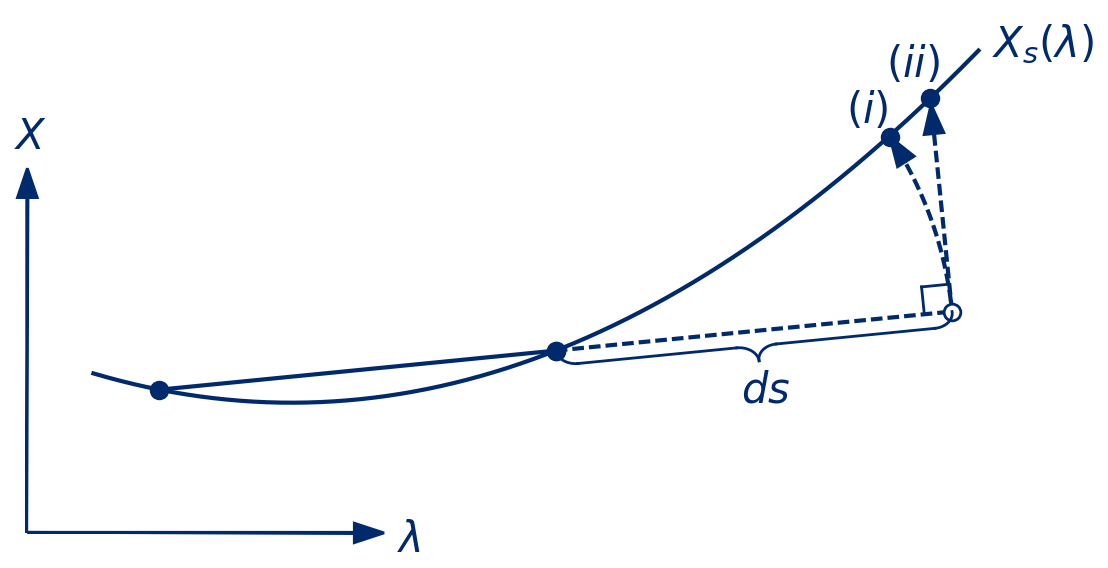}
  \caption{Sketch illustrating the basis of numerical continuation. Shown are schemes consisting of a secant predictor and (i) pseudoarclength corrector and (ii) orthogonal corrector. The secant predictor uses two  previous points on the solution branch to obtain an initial guess for the corrector via a linear extrapolation by an arclength $ds$. The pseudoarclength corrector employs Newton steps on a circle of radius $ds$ around the previous point to obtain the next point on the branch. The orthogonal corrector searches on a line perpendicular to the direction of the secant.\label{pseudoarclength}}
\end{figure}

To follow a solution branch through a fold, one may instead use the pseudoarclength method \cite{Doedel1}. In this method, a branch is not tracked using the control parameter $\lambda$ as the main continuation parameter. Instead, an intrinsic and unique property of the bifurcation curve, the arclength $s$ along the branch, is used. In particular, one employs a secant predictor advancing an arclength $ds$. Then, the predictor consists of Newton steps at fixed arclength, i.e., the true position of the next point on the branch is searched along a circle with radius $ds$ around the previous point, which is sketched as option (i) in Fig.~\ref{pseudoarclength}. That is, the next point is not searched for at fixed $\lambda$ but at fixed $s$, while $\lambda$ is adapted in the Newton steps and its exact value becomes part of the solution. With the resulting predictor-corrector scheme, it is possible to follow a branch through a saddle-node bifurcation.

Another alternative is to make the correction along a line which is perpendicular to the direction defined by the secant predictor, as sketched as option (ii) in Fig.~\ref{pseudoarclength}. We refer to this scheme a secant predictor with an orthogonal corrector and use it below in our continuation algorithm. It is simpler to realize than the pseudoarclength predictor-corrector method and approximates it with a negligible error.

We parametrize the perpendicular line on which the correction is done with a real number $u$ such that the line is given by the set of points $(\lambda_u, X_u)$. We search for a zero of the function $G_{\lambda}$ along this line. That is, we search for the zero of the real-valued function $\widetilde{G}(u) := G_{\lambda_u}(X_u)$.

If we denote a stable steady state as $X^{\ast}$, then when the state of the system $X < X^{\ast}$, we have $X < F_{\lambda}(X)$ and so $G_{\lambda}(X)$ is positive. On the other hand, when $X > X^{\ast}$, then $G_{\lambda}(X)$ is negative. Hence, near the steady state, the function $G_{\lambda}$ has a negative slope. Near an unstable steady state, it has a positive slope.

We cannot transfer this idea to the function $\widetilde{G}$ directly, because, on moving along a curved solution branch, the perpendicular line can have many different orientations. Only for a horizontal segment of a branch and a vertical perpendicular line can the above consideration be directly applied to the function $\widetilde{G}$. Along curved branches the meaning of the sign of the slope of the function $\widetilde{G}$ may change.\footnote{A straight line can be parametrized into two different directions. So, to be specific, the relation between the sign of the slope of the function $G_{\lambda}$ and the stability of the steady state depends on the direction of the parametrization. For one direction, we have stability for a positive slope, while for the other direction stability corresponds to the negative slope.} Nevertheless, via the orientation of the perpendicular line, there is a connection between this sign and the stability of the solutions on the branch being continued. So, if a change in the sign of $\widetilde{G}$ is detected during the continuation, then one has to be aware that a bifurcation point may have been passed and consider the possibility of other branches existing in the bifurcation diagram. This is discussed further in the outlook in Sec.~\ref{sec:conclusion_outlook}. 

The numerical continuation schemes just described form a well-developed method to calculate bifurcation diagrams for deterministic systems formulated as ODEs, PDEs, or as iterative maps. Many different software packages exist which implement this method, like \textsc{auto-07p} \cite{AUTO, Doedel1, Doedel2} or \textsc{pde2path} \cite{pde2path}. Furthermore, these contain many additional helpful and sophisticated tools that enable the detection of many types of bifurcations, branch switching to steady and time-periodic states, two-parameter continuation of loci of bifurcation points, like saddle-node, Hopf, and period-doubling bifurcations. This makes it possible to efficiently determine complex bifurcation diagrams.

\subsection{Stochastic continuation}
\label{sc_sec}

The continuation methods described in the previous subsection are well suited to investigate deterministic systems. We now explain how the basic concepts can be applied to stochastic systems. For such systems we must distinguish between the macroscopic and the microscopic descriptions and how these are related. In general, we use the terms ``macroscopic quantity'' and ``macroscopic dynamics'' for the quantity and dynamics we are interested in on the level of a macroscopic or ensemble-averaged description. These are the quantities that characterize the steady states in the context of bifurcation diagrams. By ``microscopic quantity'' and ``microscopic state'' we refer to individual representations of the stochastic system. Note that the terms ``microscopic'' and ``macroscopic'' may not refer literally to length scales in the system. For instance, the macroscopic state can be the mean value of a function of the microscopic state (e.g.,~the mean magnetization of the spin lattice in the Ising model) or the mean over an ensemble of microscopic states (e.g., a time average of a quantity described by a Langevin equation).

In the numerical continuation for deterministic systems described above, an equation defining the macroscopic dynamics is explicitly given. However, for a stochastic system the macroscopic dynamics is given only implicitly and the connection between the microscopic and the macroscopic level is often (but not always) of a statistical nature, since it arises from the stochastic dynamics of the microscopic states. Numerical continuation for stochastic systems, referred to as ``stochastic continuation,'' continues the steady states of the macroscopic quantities and determines their bifurcation diagrams without explicit knowledge of the time evolution of the macroscopic state. Note that by ``steady states'' of a stochastic system, we mean that the macroscopic quantity obtained by averaging is approximately steady. While microscopic states are not steady in a stochastic system, the solution measure or order parameter used to distinguish different types of microscopic states is always obtained by some averaging procedure (e.g., the mean orientation of all spins located on the Ising lattice at a particular time). Because in practice systems are always of finite size, the macroscopic quantity still slightly fluctuates. This important issue we will further discuss in Sec.~\ref{sec:difficulties}.

\begin{figure}
\includegraphics[width=0.9\hsize]{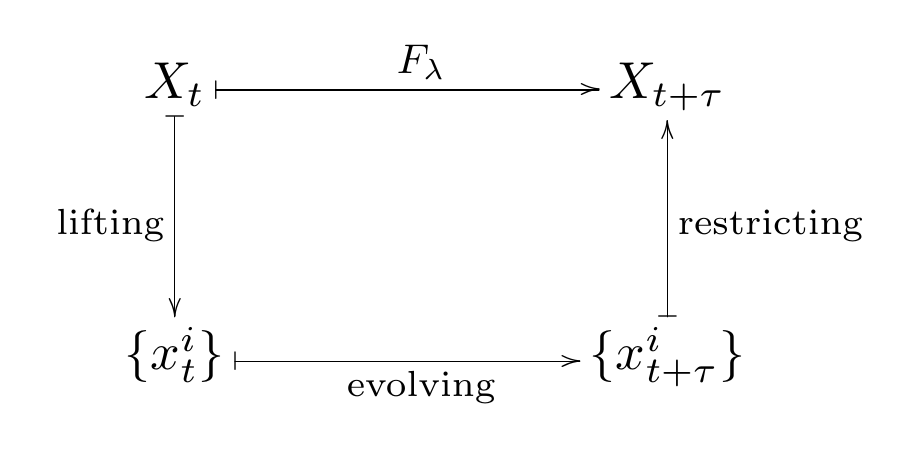}
\caption{Scheme indicating how the lifting, evolving, and restricting steps are used to evaluate a function call of $F_{\lambda}$ in the method of stochastic continuation.\label{bypass}}
\end{figure}

In principle, numerical continuation can be done in all cases where one has access to the function $F_{\lambda}$. For a stochastic system this function is not explicitly given, but one can instead use the microscopic time-stepper; see Refs.~\cite{ThQK2000pnasusa, Kev2009,ConsumerLockIn, Kueh2012sjsc, ThLS2016pa}. To do this we have to determine a way to create for each value of the macroscopic quantity $X$ a corresponding microscopic state $x$ or an ensemble of such states $\{x^i\}$. Using such a mapping, one is able to bypass the problem of not having an explicit form for the function $F_{\lambda}$. To obtain the value of the function $F_{\lambda}$ at $X_t$, we apply the following three steps: First, a corresponding microscopic state $x_t$ or a corresponding ensemble $\{x_t^i\}$ is created in the so-called \textit{lifting} step. Second, the time evolution is advanced a time interval $\tau$ employing the explicitly known microscopic dynamics to obtain the state $x_{t+\tau}$ or the ensemble $\{x_{t+\tau}^i\}$. This is the \textit{evolving} step. Finally, in the \textit{restricting} step, one returns to the macroscopic level by evaluating $X_{t+\tau}$ which is identical to $F_{\lambda}(X_t)$. This sequence of steps is shown schematically in Fig.~\ref{bypass} and bypasses the task of evaluating the unknown function $F_{\lambda}(X_t)$. Thus, every time the numerical continuation algorithm calls the macroscopic function $F_{\lambda}$ in the procedure to find the roots of the function $G_{\lambda}$ or in any other situation, the call of $F_{\lambda}$ is replaced by the three-step bypass illustrated in Fig.~\ref{bypass}.

Note that the algorithm just described includes two important additional parameters. The first is the time interval $\tau$ of the microscopic time evolution. The second is the size $M$ of the microscopic ensemble. Rules for the choice of these are discussed below in Sec.~\ref{sec:linfit} and Sec.~\ref{sec:adaptive}, respectively.

Another important issue relates to the specific definition of the lifting map that is used. This depends on the specific example being considered and it's microscopic dynamics. We present some general ideas how to do this in Sec.~\ref{sec:strulift} below.

A further issue for stochastic systems relates to the fact that $\widetilde{G}$ is also a fluctuating quantity, i.e., to find a zero of $\widetilde{G}$ one must deal with the statistical fluctuations. Here we use the $\text{C}^3\text{R}$ method introduced in Ref.~\cite{ThLS2016pa} to prevent the occurrence of outliers which can lead to poor continuation results. In this method, every approximation which lies outside of a suitably defined interval around the initial guess is repeated a predefined number of times. This greatly reduces the probability of outliers from statistical fluctuations leading to poor results. This is used in combination with the pseudoadaptive parameter adjustment introduced in the next section.

\subsection{Challenges of stochastic continuation -- Towards an effective procedure}
\label{sec:difficulties}

The stochastic continuation approach presented in the previous section faces a number of important challenges. Here, we discuss these and give methods to remedy the difficulties.

\subsubsection{Fluctuations in $\widetilde{G}$ -- Using linear fits}\label{sec:linfit}

When dealing with stochastic systems, the system size and the ensemble size that can be simulated are always finite. As a consequence, the function $\widetilde{G}$ exhibits statistical fluctuations which complicate the root finding and induces uncertainties.

\begin{figure}
  \includegraphics[width=0.9\hsize]{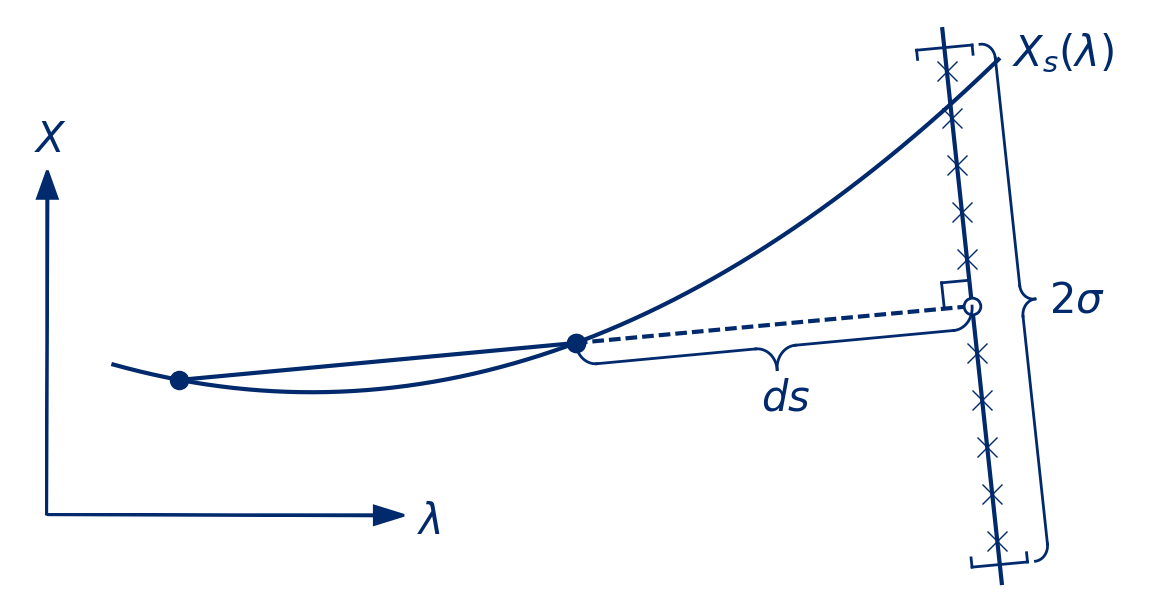}
  \caption{To determine the next point on the branch of steady solutions, the function $\widetilde{G}$ is evaluated at several points within an interval around the initial guess (indicated by small crosses) on the line perpendicular to the line defined by the secant predictor. The length of the interval is controlled by the parameter $\sigma$ that is chosen in such a way that the function $\widetilde{G}$ is sufficiently linear and that the required root is contained in it.\label{fit_geom}}
\end{figure}

\begin{figure}
  \includegraphics[width=0.9\hsize]{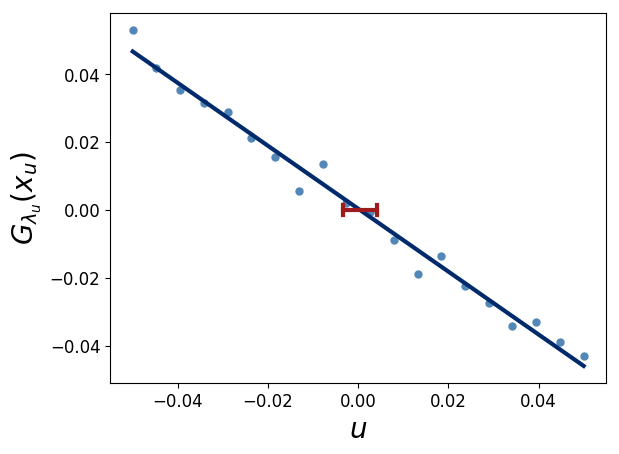}
  \caption{The plot shows a typical point cloud obtained through evaluations  of the function $\widetilde{G}(u) = G_{\lambda_u}(X_u)$ on an interval as illustrated in Fig.~\ref{fit_geom} with $\sigma=0.05$. The point cloud is shown together with the resulting linear fit used to determine the zero of the fluctuating function. The standard deviation of the horizontal distances of all points to the fit line is indicated as a horizontal red bar. It serves as a measure for the statistical error in the zero. In this particular example, the data are from stochastic continuation of the branch of stable steady states of the Ising model at the reduced temperature $T'=3.1$. Here $T'$ takes the role of $\lambda$. The size of the spin lattice is $100\times 100$ while the parameters of the continuation algorithm are $\tau=496$, $M=39$, $\sigma=0.05$, $ds = 0.1$, and $N_{\text{fit}} = 20$. The values of $\tau$, $M$, and $\sigma$ are obtained by our adaptive parameter choice (as explained in Sec.~\ref{sec:adaptive}) with $\Delta_d = 0.005$ and $\theta_d = 1.0$. The slope of the fit is $\theta = 0.93$ and the error in the zero is $\Delta = 0.0037$. \label{point_cloud_fit}}
\end{figure}

Here we solve this problem by employing linear fits. Namely we evaluate a finite number $N_{\text{fit}}$ of function values of $\widetilde{G}$ equidistantly spaced on the line perpendicular to the secant prediction that defines the correction step in the orthogonal corrector -- see Sec.~\ref{sec:numcont}. We use here $10\leq N_{\text{fit}}\leq50$. This is done up to a suitable distance from the initial guess, as illustrated in Fig.~\ref{fit_geom}. This yields the function $\widetilde{G}$ evaluated in the form of a cloud of points. The cloud is then approximated by a linear best fit whose root is then employed as the approximation for the zero of $\widetilde{G}$. A typical example is displayed in Fig.~\ref{point_cloud_fit}. This procedure is very robust if an appropriate range for the evaluation of $\widetilde{G}$ and the fitting is defined. Of course, the function $\widetilde{G}$ should be sufficiently linear over this range. Thus, on the one hand, the range considered should be sufficiently small for this to be true, but on the other hand, the root should lie within this range, because an extrapolation beyond the range considered can lead to poorly controlled uncertainty. As a result, we must choose a step size for the prediction step to be sufficiently small so that the initial guesses for the correction and the zero of $\widetilde{G}$ are sufficiently close together. We define a parameter $\sigma$ that specifies the range to be considered, as indicated in Fig.~\ref{fit_geom}. A good value turns out to be $\sigma=0.5ds$, as we show next.

The fitting procedure works in a stable way because it moderates the fluctuations of the function $\widetilde{G}$. Additionally, it has the advantage of allowing us to estimate the size of the statistical error $\pm\Delta$ in the calculated zero. As a measure for $\Delta$ we use the standard deviation of the horizontal distances of all the calculated points to the fitted line -- see Fig.~\ref{point_cloud_fit}. With $N$ evaluated values of the function $\widetilde{G}$ with the abscissae $u_1,...,u_N$ and the ordinates $\tilde{g}_1,...,\tilde{g}_N$ and the linear fit $\widetilde{G}=\theta u + b$, we obtain
\begin{equation}\label{error_measure}
	\Delta := \sqrt{ \frac{1}{N} \sum_{i=1}^N \left(\frac{\tilde{g}_i-b}{\theta} - u_i \right)^2 }.
\end{equation}
Without our fitting technique, simple numerical root finding methods generally fail, because of the fluctuations in $\widetilde{G}$. Independently of the particular fitting technique, it is possible to determine the zeros by applying more advanced root finding methods which take derivatives of the considered function into account \cite{ThLS2016pa}.

\subsubsection{Divergence of timescales -- Adaptive choice of numerical parameters}\label{sec:adaptive}

\begin{figure}
  \includegraphics[width=0.9\hsize]{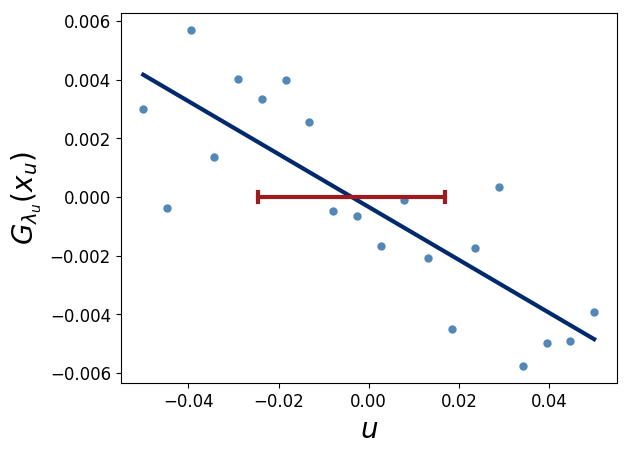}
  \caption{Comparison of this figure and Fig.~\ref{point_cloud_fit} shows the influence of the size of the time span $\tau$ used for the microscopic time evolution. Here the point cloud is obtained by evaluating the function $\widetilde{G}(u) = G_{\lambda_u}(X_u)$ with $\tau=10$, which is too small. All the other parameters are as in Fig.~\ref{point_cloud_fit}. The slope of the straight-line fit is $\theta = 0.09$, the error in the zero is $\Delta = 0.0208$.\label{lowtau}}
\end{figure}

Roughly speaking, one can view the dynamics of a typical system to be considered as being composed of a deterministic part and a stochastic part. The deterministic part controls the slope $\theta$ of the function $\widetilde{G}$, while the stochastic part causes it to fluctuate. The larger the influence of the deterministic part, the larger is the absolute value of the slope $|\theta|$. For small $|\theta|$, fluctuations dominate and any information we have about the location of the zero is not reliable (cf.\ the case in Fig.~\ref{lowtau}). For large $|\theta|$, the uncertainties are small (cf.\ the case in Fig.~\ref{point_cloud_fit}).

It is well known that on approaching a bifurcation point (sometimes called a ``tipping point'' \cite{DWCD2014ccp}), the typical timescales of the system dynamics diverge -- often referred to as ``critical slowing down.''
This implies that if a fixed time span $\tau$ is used for the microscopic time evolution, the slope of $\widetilde{G}$ tends to zero when approaching the bifurcation point. This causes a significant loss of precision for these parameter values, which is unfortunate, because this concerns the (generally) most important details of a bifurcation diagram.

Our remedy for this issue consists of dynamically adjusting the value of $\tau$ to the timescale typical for the system at any given point in the bifurcation diagram. In other words, when the dynamics of the macroscopic observable slows down the microscopic time evolution is prolonged by adapting $\tau$.  For example, Fig.~\ref{lowtau} displays results for $\widetilde{G}$ with $\tau=10$, which is too small while Fig.~\ref{point_cloud_fit} shows the corresponding result obtained by adapting $\tau$ to 496.

Based on our accumulated experience, it is possible to define suitable values of $\tau$ ``by hand''. We call this approach a \textit{pseudoadaptive} continuation. We have done this for the first continuation results of the Ising example (cf.~Sec.~\ref{sec:ising}). Combined with the $\text{C}^3\text{R}$ method and the lifting procedure presented in the next section, very good results can be obtained. However, the pseudoadaptive method is very labor intensive and depends greatly on the user's experience. This can be avoided by using the fully adaptive adjustment of the parameter $\tau$ discussed next.

\begin{figure}
  \includegraphics[width=0.9\hsize]{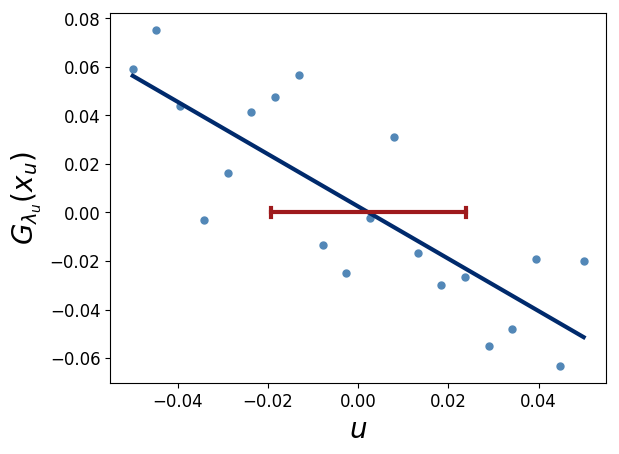}
  \caption{Comparison of this figure and Fig.~\ref{point_cloud_fit} shows the influence of the size of the microscopic ensemble $M$ used for the microscopic time evolution. Here the point cloud is obtained by evaluating $\widetilde{G}(u) = G_{\lambda_u}(X_u)$ with $M=2$, which is too small. All the other parameters are as in Fig.~\ref{point_cloud_fit}. The slope of the fit is $\theta = 1.08$, the error in the zero is $\Delta = 0.0216$.\label{lowM}}
\end{figure}

Furthermore, due to the law of large numbers, the size of the microscopic ensemble $M$ also has an influence on the magnitude of fluctuations in $\widetilde{G}$ and the precision of the continuation results. For example, Fig.~\ref{lowM} shows $\widetilde{G}$ evaluated for $M=2$, which is too small. This should be compared with the corresponding results obtained by adapting $M$ to 39 displayed in  Fig.~\ref{point_cloud_fit}. This comparison illustrates that the parameters $\tau$ and $M$ greatly influence the exactness of the method. They also have a large effect on the numerical cost, since the ensemble of microscopic time evolutions usually forms the largest contribution to the computation time. Thus, these parameters should be chosen as small as possible and as large as needed. Note also that the parameter $\sigma$ discussed in Sec.~\ref{sec:linfit} is important in this context.

Our proposed adaptive adjustment takes the three parameters $\tau$, $M$, and $\sigma$ into account. For every parameter, we specify a condition which should be fulfilled to effect a reliable determination of the root of the function $\widetilde{G}$ and we implement a rule for modifying the parameter values when these conditions are not fulfilled. For the parameters $\tau$ and $M$ we also define upper limits ($\tau_{\text{max}}$ and $M_{\text{max}}$) that reflect the available computing resources. Since these rules are based on estimates, they are iteratively applied until all of the conditions are met. As initial values for the parameters, we use the adjusted parameter values from the previous continuation step. To make the adjustment of $\tau$ more efficient, we initially extrapolate the change between the previous two continuation steps. Namely if the adaptive adjustments in the previously performed continuation steps produced the series of values $\tau_0, \tau_1, ..., \tau_{n-1}$, then we use
\begin{equation}
	\tau_n^{\text{guess}} := \tau_{n-1} + (\tau_{n-1} - \tau_{n-2})
\end{equation}
as initial value for $\tau$.

\begin{figure*}
\includegraphics[width=0.9\textwidth]{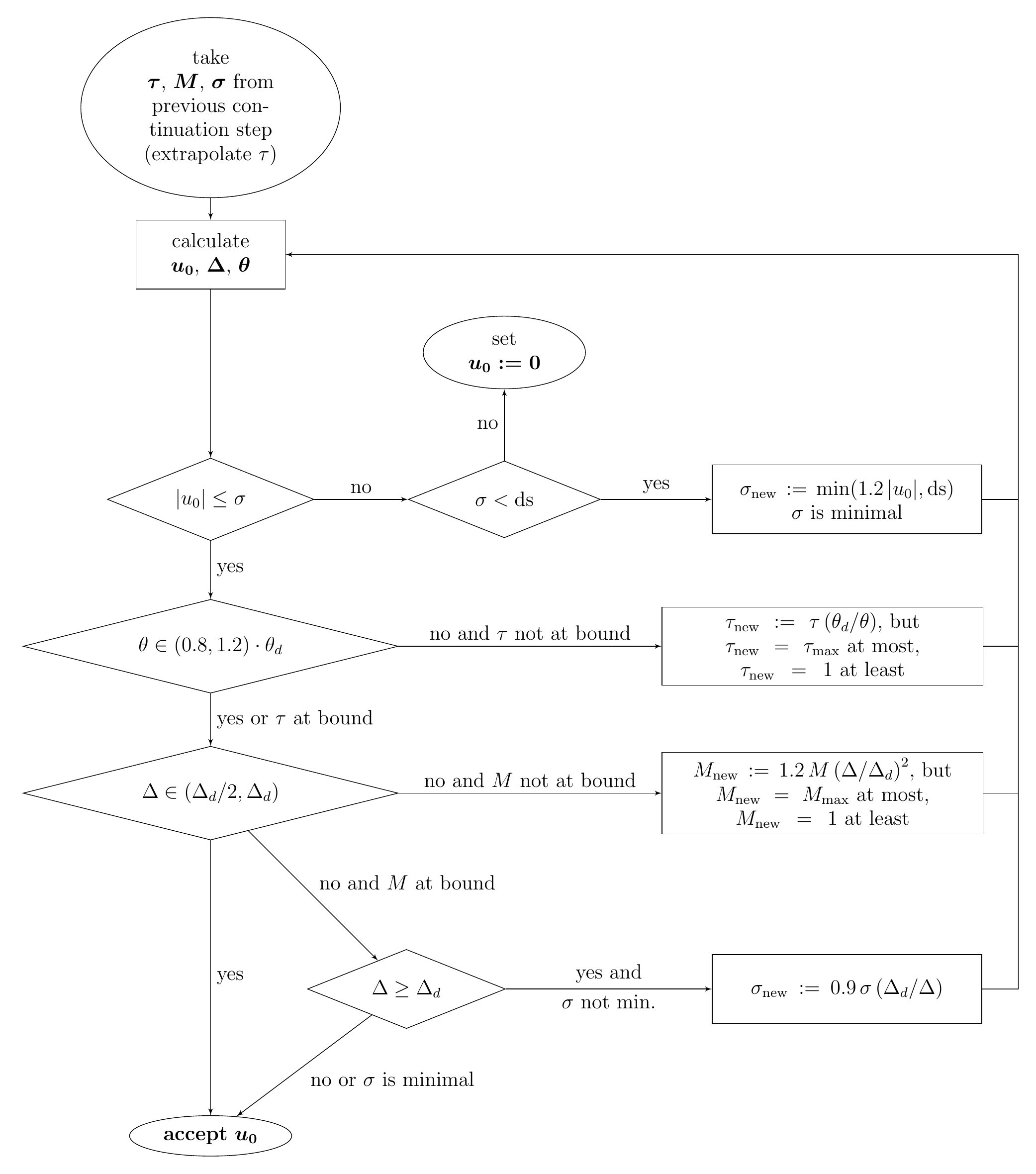}%
\caption{Flow chart for the algorithm employed to adaptively adjust the parameters $\tau$, $M$, and $\sigma$. In each continuation step, the adjustments are applied iteratively until every condition is fulfilled. \label{flowChart}}
\end{figure*}

It is useful to start the continuation procedure with low values of $\tau$ and $M$ in a region of the bifurcation diagram where no complications are expected, e.g., on a stable branch far away from bifurcation points, and to let the adaptive control work while approaching bifurcation points. Figure~\ref{flowChart} gives a flow chart that illustrates how our algorithm adjusts the various parameters. This is discussed below in the order of parameter update.

Most importantly, the interval $\sigma$ in which we fit $\widetilde{G}(u)$ must be appropriate for a linear approximation. Furthermore, the zero should lie within the interval, because extrapolation outside the interval leads to uncontrolled errors. Thus, if the absolute value of the calculated zero $u_0$ is greater than $\sigma$, then we set $\sigma=1.2 |u_0|$. However, $\sigma$ should not be greater than the continuation step size $ds$. If this condition cannot be fulfilled, then we instead decrease the value of $ds$. If this is not possible and $|u_0|$ remains greater than $\sigma$ despite all adjustments of the other parameters, then we define the root to be zero. That is, we do not change the direction for the next step along the branch of the bifurcation diagram. This can also occur close to a bifurcation point, since in this situation a reliable result cannot be achieved with an acceptable numerical effort. In this case, one blind step in the previous direction is the best option.

To adjust the parameter $\tau$, we use the slope $\theta$ of the linear fit. This slope should be in the interval $(0.8,1.2) \theta_d$, where $\theta_d$ is the desired slope. Based on experience, a value of $\theta_d=1$ leads to good results. However, it may be reasonable to choose a slightly lower value (down to $\theta_d=0.5$) to reduce the numerical cost.

For $\tau$ and $\theta$ we assume a linear relation as the first estimate, i.e.,\ if $\theta$ does not lie in the above interval, we set
\begin{equation}
	\tau_{\text{new}} := \tau \frac{\theta_d}{\theta}.
\end{equation}
To adapt the parameter $M$, we take the error $\Delta$ into account. We determine a maximal error (denoted by $\Delta_d$) and change $M$ if $\Delta$ lies outside of the interval $(\Delta_d/2, \Delta_d)$. If one considers $N$ independent and identically distributed random variables, then the standard deviation of their mean is the standard deviation of one variable times $N^{-1/2}$. Thus, we suppose a linear relation between $\Delta$ and $M^{-1/2}$. Including a safety margin, we use
\begin{equation}
	M_{\text{new}} = 1.2M \left( \frac{\Delta}{\Delta_d} \right)^2.
\end{equation}
If the error $\Delta$ is still greater than $\Delta_d$ after the adjustment of $M$, then we decrease the parameter $\sigma$ using
\begin{equation}
	\sigma_{\text{new}} = 0.9\sigma \frac{\Delta_d}{\Delta}.
\end{equation}
For the parameters $\tau$ and $M$, it can be useful to set their minimum values to 10, for example. Otherwise, a small adjustment by a factor close to 1 has no effect, since $\tau$ and $M$ are natural numbers.
\subsubsection{Lifting is one-to-many mapping -- Structure lifting}\label{sec:strulift}
\begin{figure}
  \includegraphics[width=0.9\hsize]{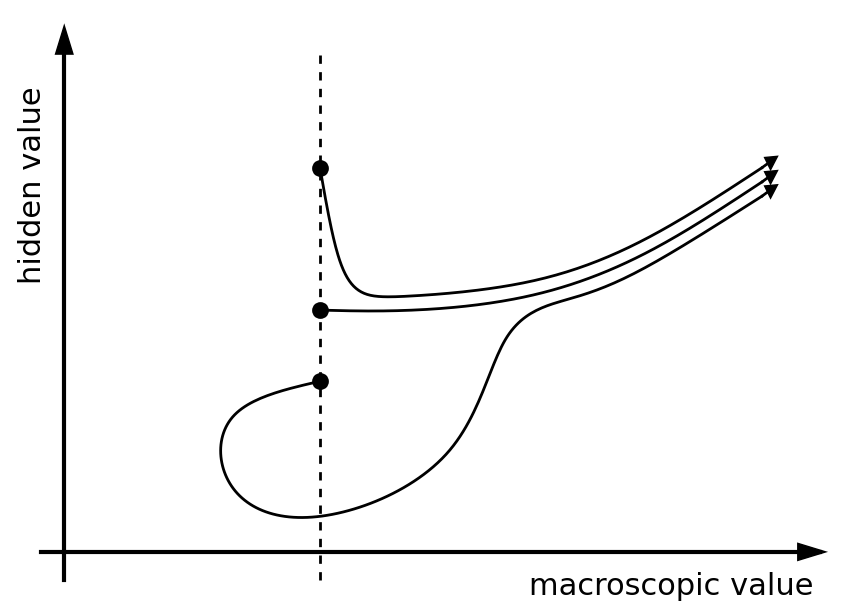}
  \caption{A sketch illustrating some potential problems with the lifting procedure due to it being a one-to-many mapping. The different possibilities to define a microscopic state belonging to one value of the macroscopic quantity are indicated by the axis with the label ``hidden value.'' The lines depict different trajectories in phase space and show that large excursions are possible before randomness is ``healed out.''\label{liftingOTMM}}
\end{figure}

The final issue to discuss concerns the lifting procedure needed to initiate microscopic states from a known macroscopic state (see Fig.~\ref{bypass}). Obviously, lifting is not unique but corresponds to a one-to-many mapping. This gives a large procedural freedom and requires a detailed discussion of how the lifting is done. In fact, the structure of the microscopic state can influence the resulting dynamics of the macroscopic observable, as illustrated in Fig.~\ref{liftingOTMM}. In the best case, errors from an inconvenient lifting procedure are healed over a small initial interval within the time span $\tau$ of the evolving step. In the worst case, lifting errors crucially impact the macroscopic dynamics leading to wrong results in the root finding process. Eventually, this also leads to uncertainties in the choice of the parameter $\tau$. Further, for reasons of computational efficiency, this parameter has to be chosen as small as possible. Consequently, the quality and precision of the evolving step strongly depend on having an effective lifting procedure.

In fact, the ambiguity of the lifting procedure is potentially the most dangerous problem. Often, microscopic (fluctuating) steady states exhibit specific spontaneously formed structures which are crucial to the time evolution of the macroscopic observable. However, their formation from unstructured microscopic states may take a time span that is too large to be able to accommodate it within the chosen $\tau$. This problem can be tackled if one bases the stochastic continuation on something more refined than simple time simulation from unstructured initial states.

We explain our approach to solve this problem employing our first example, the Ising model. There, the microscopic state corresponds to a particular state of a lattice of spins $\{ s_i \}$ while the corresponding macroscopic observable is magnetization $m = \langle s_i\rangle$ obtained by averaging over the microscopic state. When the magnetization is near zero and the temperature is near or below $T_C$, a typical spin configuration exhibits a clustered structure, which has an influence on the dynamics of the magnetization (examples can be seen below in Sec.~\ref{sec:ising}).

\begin{figure}
  \includegraphics[width=0.9\hsize]{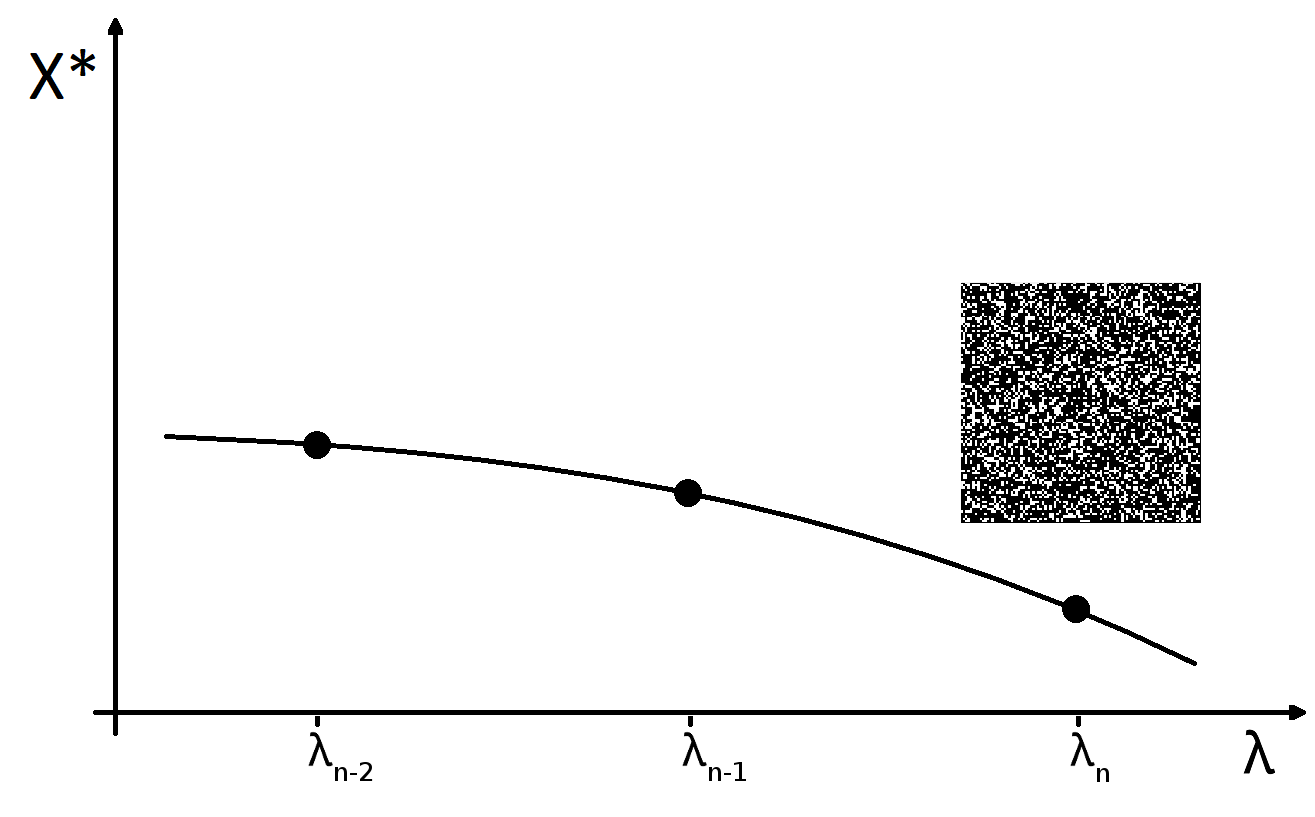}
  \caption{Scheme of the standard random lifting procedure (also called structure lifting of 0th order): In continuation step $n$ the microscopic dynamics is initiated with a random state that corresponds to the predicted value of the macroscopic observable $X^\star$.
\label{randomLifting}}
\end{figure}

 A naive mapping in the lifting procedure is to create a spin lattice with the orientations randomly chosen to give the correct magnetization. However, this leads to spin orientations without spatial correlations. We call this \textit{random lifting} and illustrate it in Fig.~\ref{randomLifting}. During the time evolution of the evolving step, this disordered structure starts to form clusters. However, the time span $\tau$ is not sufficiently large to complete this process. Furthermore, the correlation information is forgotten after the microscopic time evolution is terminated and in the next application of the lifting, a disordered structure is again produced.

\begin{figure}
  \includegraphics[width=0.9\hsize]{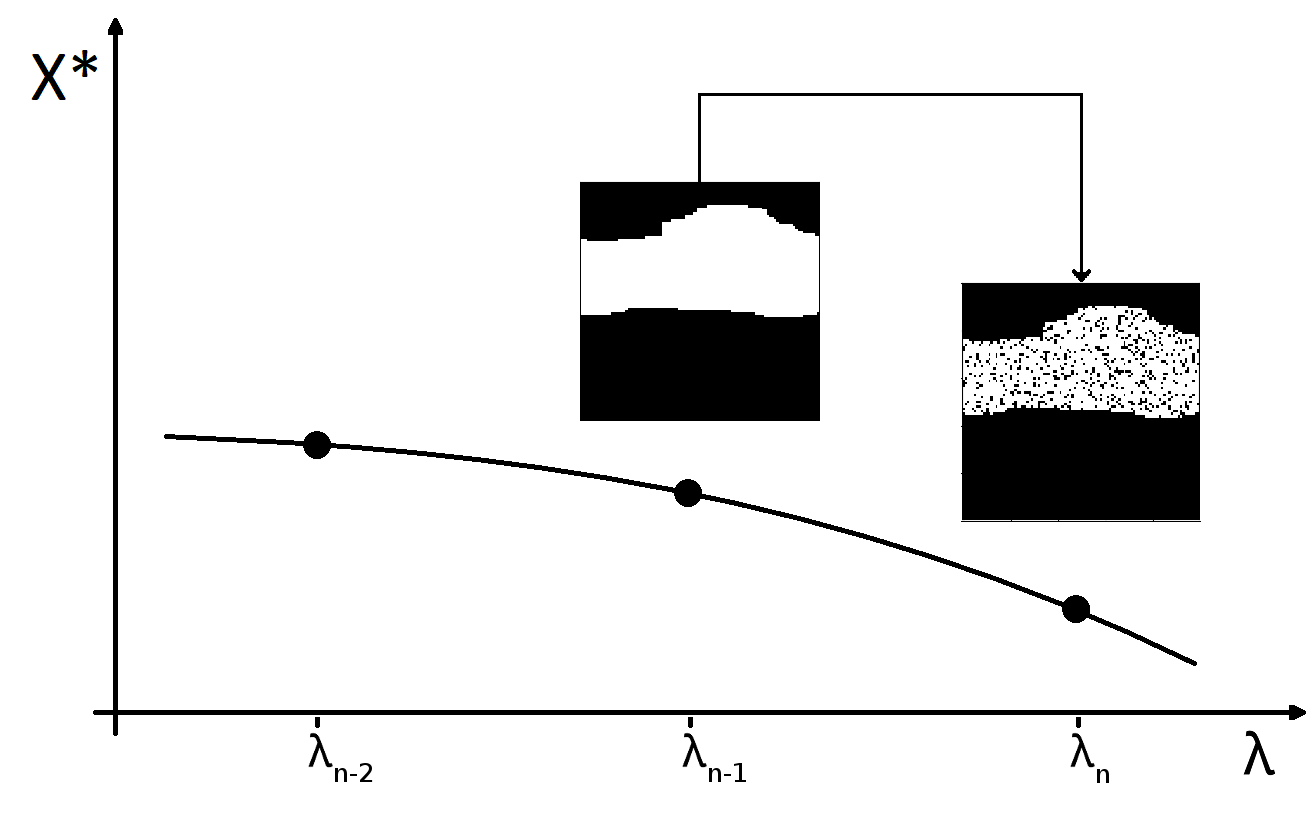}
  \caption{Scheme of the structure lifting of first order: In continuation step $n$ the microscopic dynamics is initiated with a structured microscopic state obtained by adapting a final state of the microscopic dynamics from the previous continuation step $n-1$. The adaptation is done randomly and brings the macroscopic observable $X^\star$ to the value requested by the prediction step.
\label{structureLifting}}
\end{figure}

Instead, we introduce \textit{structure lifting}, a lifting technique that takes information from previous microscopic time evolutions into account. In particular, the final states of the evolving process are kept, which include all the information concerning their internal structure. Then these states are employed in the lifting procedure at the next value of the control parameter. Because successive applications of the lifting step are always done at a value of the macroscopic observable that does not strongly differ from the previous value, only minor adjustments of the microscopic states have to be done. In the simplest case, these can be realized in a spatially random way, as illustrated in Fig.~\ref{structureLifting}. This particular procedure we call \textit{first-order structure lifting}.

\begin{figure}
  \includegraphics[width=0.9\hsize]{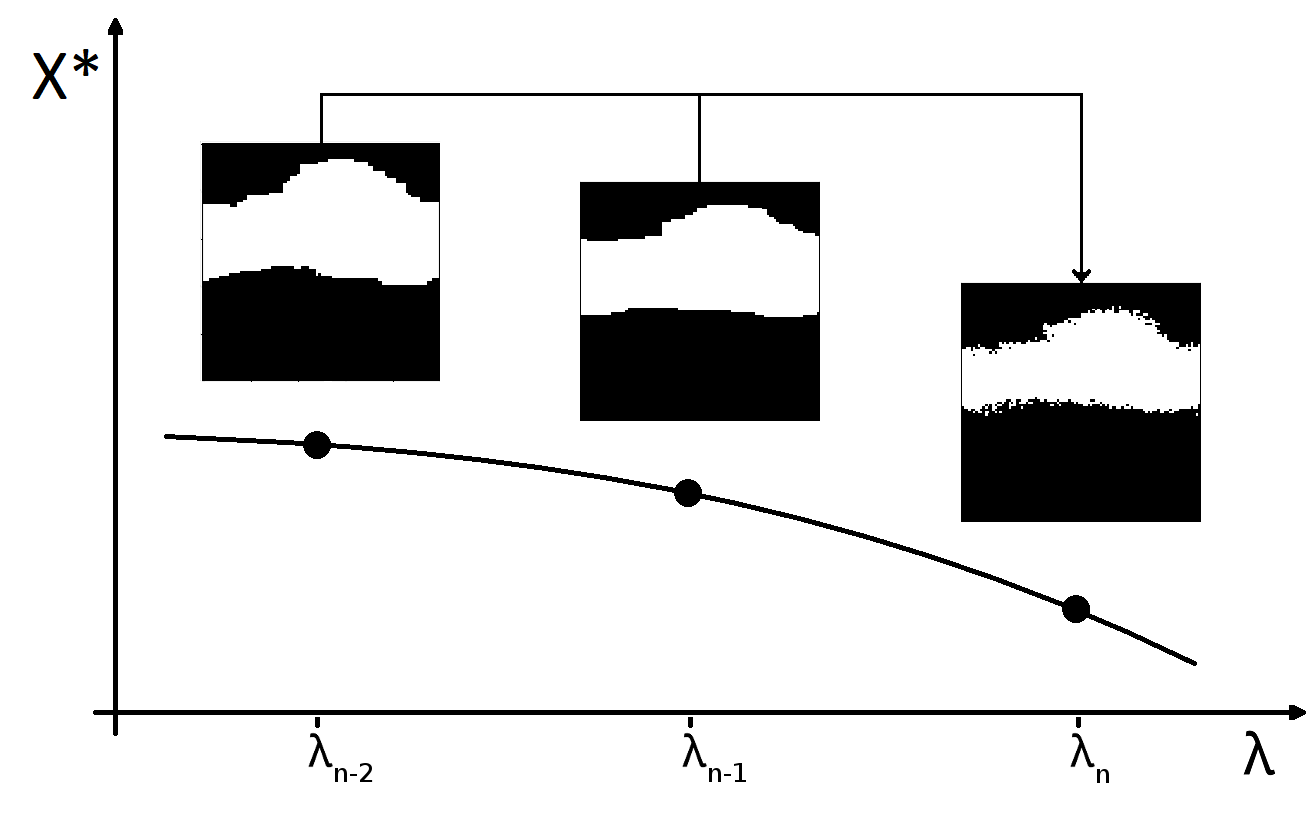}
  \caption{Scheme of the structure lifting of second order: In continuation step $n$ the microscopic dynamics is initiated with a structured microscopic state obtained using information from the final states of the microscopic dynamics from the two previous continuation steps $n-2$ and $n-1$. This allows one to adapt the microscopic state mainly in regions where most changes occurred between steps $n-2$ and $n-1$.
    \label{2ndstructureLifting}}
\end{figure}

To further improve the efficiency of the scheme, it is possible to go one step further and perform a \textit{second-order structure lifting}. Here, information is kept concerning microscopic structures that have resulted from evolving steps in the two previous continuation steps. This allows one to extract information about the typical changes of the structure along the already calculated part of the solution branch. This information is then used to form the initial microscopic state in the next lifting procedure in a more precise way, as illustrated in Fig.~\ref{2ndstructureLifting}. Namely the necessary random changes are concentrated in regions where most changes occurred between the two previous steps. In the example of the Ising model, this means that the minor adjustments are not uniformly distributed in space, as in the first-order structure lifting. Now, with the highest probability they are located at the interfaces between the clusters. We achieve this in practice by employing a Gaussian kernel to increase the probability density of changes around the locations where the structure changed previously.

In principle, this concept for the lifting procedure can be extended to \textit{nth-order structure lifting}. This method produces naturally evolved microscopic states without the need for long relaxation times in every evolving step. Of course, to employ structure lifting at the beginning one needs to provide the algorithm with one or more microscopic states. Depending on the specific situation, these can be generated using random lifting or a long relaxation.

Next, we illustrate the application of the proposed amendments to stochastic continuation via three examples. The first is the Ising model in Sec.~\ref{sec:ising}, which we have touched on already; an active Ising model in Sec.~\ref{sec:activeIsing}; and a stochastic Swift-Hohenberg equation in Sec.~\ref{sec:SwiftHohenberg}.
\section{Ising model\label{sec:ising}}
The Ising model Hamiltonian is given above in Eq.~\eqref{eq:Ising_H}. As mentioned there, the macroscopic observable is the magnetization $m := \langle s_i \rangle$. Depending on the dimensionless temperature $T':=k_\mathrm{B}T/J$, different stable and unstable states exist. For the 2D Ising model, an exact solution in terms of the free energy is given by Onsager \cite{Onsa1944pr}, while Yang first derived the formula for the spontaneous magnetization \cite{Yang1952} (also announced by Onsager at a conference in 1949 \cite{baxter2016exactly}). As the exact result is available, this example provides a valuable test for the precision of our approach.

\begin{figure}
  \includegraphics[width=0.9\hsize]{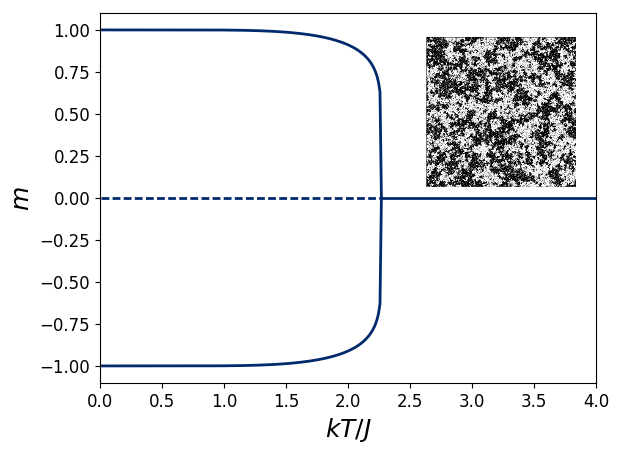}
  \caption{The exact (Onsager) bifurcation diagram of the two-dimensional Ising model. The solid and dashed lines indicate the branches of stable and unstable states, respectively. The inset gives a typical spin configuration near the critical temperature (at $T'=2.285$) that shows spin clusters of different sizes. \label{onsager}}
\end{figure}

The exact solution is displayed in Fig.~\ref{onsager}. It shows a pitchfork-like bifurcation at the \textit{critical temperature} $T_c$, whose numerical value is given by $T'_c=k_\mathrm{B}T_c/J\approx 2.269$. At $T_c$ two branches of states of finite magnetization emerge from the trivial state of zero magnetization. For temperatures above $T_c$, the zero magnetization state is the only state, i.e., it is globally stable. For $T<T_c$, this state is unstable and the states on the two branches emerging at $T_c$ are stable. They are related by symmetry, i.e., they have the same energy. This implies that below $T_c$ the system undergoes a spontaneous symmetry breaking. The exact solution for the magnetization $m$ is given by
\begin{equation}\label{eq:exact_m}
 m = \pm\left\lbrack 1 - \left( \sinh\left(\frac{2}{T'}\right) \right)^{-4} \right\rbrack^{1/8}.
\end{equation}
For temperatures $T'<T'_c$, and for $T'\geq T'_c$ we have $m=0$. Close to $T'_c$, one write Eq.~\eqref{eq:exact_m} as
\begin{equation}
m \approx m_0\, |1-T'/T'_c|^{\beta},
\end{equation}
with the \textit{critical exponent} $\beta$. According to the exact solution, $\beta=1/8 = 0.125$.

Mean-field approximations, such as the Bragg-Williams approximation \cite{reichl1999modern,hughes2014introduction}, generally overestimate the critical temperature (e.g.,\ giving $T'_c = 4$) and give the critical exponent to be $\beta = 1/2$, much larger than the exact value \cite{baxter2016exactly}. The mean-field exponent $\beta = 1/2$ corresponds to the value expected for a standard pitchfork bifurcation \cite{Strogatz2014}.

The microscopic dynamics of the spin lattice in the evolving step is given by the Metropolis algorithm \cite{MRRT1953JCP}. That is, we  randomly choose a lattice site $i$ and then calculate the difference in the energy 
\begin{equation}
	\Delta E = 2J \sum_{\mathclap{\text{nn of }i}} s_i s_j,
\end{equation}
resulting from flipping the spin at the chosen site. This spin flip is then accepted with a probability of 
\begin{equation}\label{flipprob}
  p = \min(e^{-\Delta E/k_\mathrm{B}T}, 1).
\end{equation}
Otherwise, the flip is rejected and the spin configuration remains the same. Note that we use periodic boundary conditions. We then randomly pick another lattice site and repeat the above process. We define one time step $\tau$ as being the time to attempt $N$ spin flips, i.e.,\ one attempt per spin on the lattice.

\begin{figure}
  \includegraphics[width=0.9\hsize]{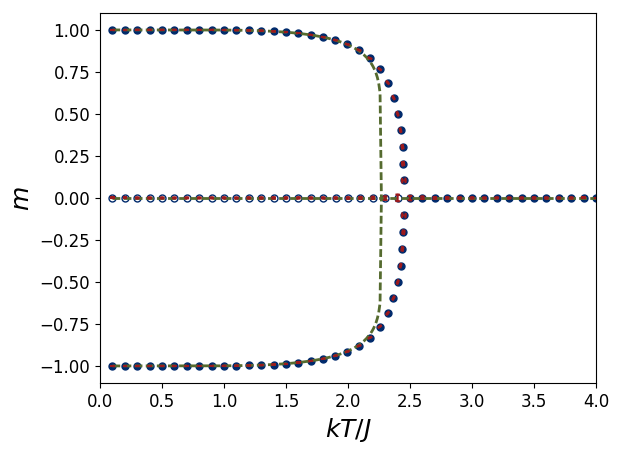}
  \caption{Bifurcation diagram of the steady states of the 2D Ising model obtained using the presented stochastic continuation algorithm in the case with random lifting. The blue circles with red error bars are the continuation results while the green dashed lines represent the exact solution. The $\text{C}^3\text{R}$ method is used and a pseudoadaptive adjustment of $\tau$. Namely $\tau$ is linearly increased from 1 to $\tau_{\text{max}}$ while approaching the bifurcation point (here, first a rough idea of the location of the bifurcation point can be obtained by a continuation run with a fixed value of $\tau$). For the nontrivial branches with $m\neq0$ we chose $\tau_{\text{max}}=70$ while for the trivial $m=0$ branch $\tau_{\text{max}}=50$ ($\tau_{\text{max}}=100$) when approaching the critical point from the left (right). The remaining parameters are $ds=0.1$, $N_{\text{fit}}=100$, $M=1$, and $\sigma=0.05$. The size of the spin lattice is $1000\times 1000$. \label{ising1}}
\end{figure}

\begin{figure}
  \includegraphics[width=0.9\hsize]{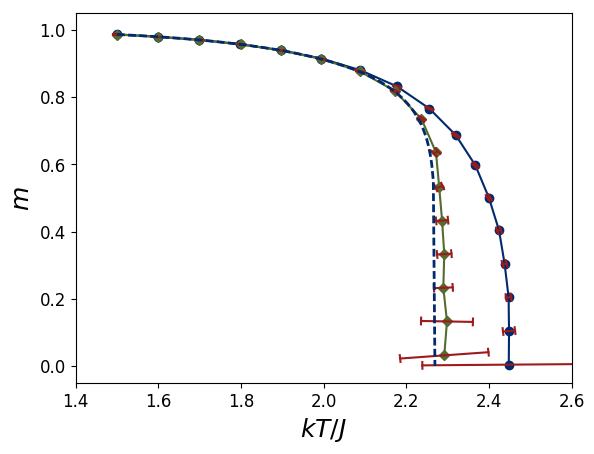}
  \caption{Bifurcation diagram showing the $m\neq0$ steady states of the 2D Ising model obtained using random lifting (blue circles) and first-order structure lifting (green diamonds). The remaining line styles, settings of the continuation algorithm and parameters are as in Fig.~\ref{ising1}. \label{ising2}}
\end{figure}

\begin{figure}
  \includegraphics[width=0.9\hsize]{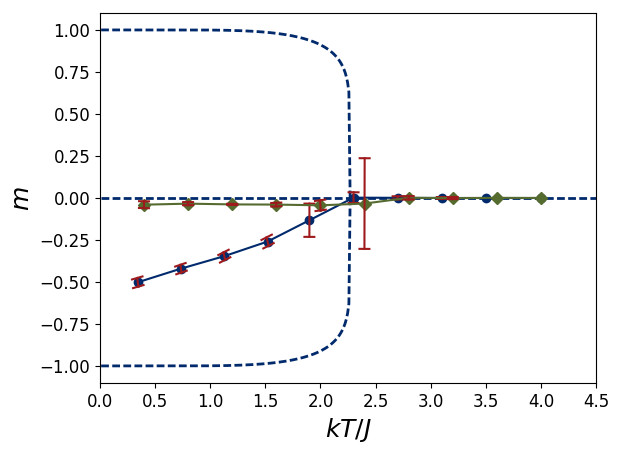}
  \caption{Bifurcation diagram of the trivial steady states of the 2D Ising model as obtained using first-order structure lifting (blue circles) and second-order structure lifting (green diamonds). The dashed lines belong to the exact solution. The fully adaptive choice of $\tau$, $M$, and $\sigma$ is used. The remaining parameters are $ds=0.4$, $N_{\text{fit}}=10$, $\Delta_d = 0.04$, and $\theta_d = 1.0$. The size of the spin lattice is $500\times 500$.\label{ising3}}
\end{figure}

We now present results obtained using the stochastic continuation algorithms described above applied to the Ising model. Those obtained using the pseudoadaptive and the adaptive parameter choice (see Sec.~\ref{sec:adaptive}) are in Figs.~\ref{ising1} and \ref{ising2}, and Figs.~\ref{ising3} and \ref{loglogneu}, respectively. Note that we treat a change in the size of the microscopic ensemble as being equivalent to changing the lattice size. When we apply the pseudoadaptive parameter choice, the lattice size remains fixed at $1000\times 1000$. When employing the adaptive parameter choice, we use the smaller lattice sizes of $100\times 100$ and $500\times 500$ because this allows for more flexible adjustment. Note that for all lattice sizes used here, finite size effects on the values of all quantities studied here are negligible. They only become important for smaller lattices, e.g., of size of $10\times 10$ or $20\times 20$ (cf.~Ref.~\cite{IsingFiniteSize}).

At first, we employ random lifting, i.e., to form the initial microscopic state with magnetization $m$, we choose the orientation of each spin according to
\begin{equation}
 s_i = \left\{ \begin{array}{r} 1,\quad\text{with probability}\quad (1+m)/2\\ -1,\quad\text{with probability}\quad (1-m)/2 \end{array}\right.
\end{equation}
This procedure produces a spatially disordered spin configuration whose magnetization is -- due to the law of large numbers -- very close to the value $m$. Stochastic continuation results in the bifurcation diagram shown in Fig.~\ref{ising1}. Overall, there is a good agreement, except for the stable $m\neq0$ branches near the critical temperature. Extrapolating these to $m=0$, we find that this continuation method gives the critical temperature to be ${T'_c}^{\text{cont}} = 2.447$. This differs from the exact value of 2.269 by about 8\%, but it is much better than the mean-field approximation result.

Employing first-order structure lifting, one can further improve this result (see Fig.~\ref{ising2}), getting an extrapolated value of the critical temperature ${T'_c}^{\text{cont}} = 2.296$, which differs only by 1.2\% from the exact value. This highlights the importance of the specific way the lifting procedure is done and how important it is to use information regarding the spatial structure of the microscopic state. It also shows the influence this has on the dynamics of the microscopic state. In the Ising example, the clustering of the spin orientations is particularly important for the dynamics near the bifurcation point. The inset of Fig.~\ref{onsager} gives a typical spin configuration with cluster formation at a temperature near $T'_c$.

However, continuation with this first-order structure lifting encounters problems along the trivial branch of zero magnetization. Namely the algorithm is not able to follow the unstable solution branch (see the blue circles for $T'<T'_c$ in Fig.~\ref{ising3}). Alternatively, employing the second-order structure lifting, which has a higher probability that adjustments are located near the borders of the clusters, result in microscopic states which are more natural and exhibits a reliable microscopic dynamics. The results obtained are shown in Fig.~\ref{ising3}. There we see that the second-order structure lifting gives a better result for the unstable part of the trivial $m=0$ branch.

This example shows that the first-order structure lifting exhibits a specific problem when applied to microscopic states that are strongly clustered. As explained above, the adjustments made to the spin configuration during the lifting process are spatially random. Thus, it is very likely that the adjustments occur at lattice sites which are in the midst of the spatially ordered regions. Then, with a high probability [see Eq.~(\ref{flipprob})], the microscopic dynamics reverses these adjustments resulting in a state with the wrong average magnetization. Hence, the continuation can be led astray. With second-order structure lifting, this issue is improved.

\begin{figure}
  \includegraphics[width=0.9\hsize]{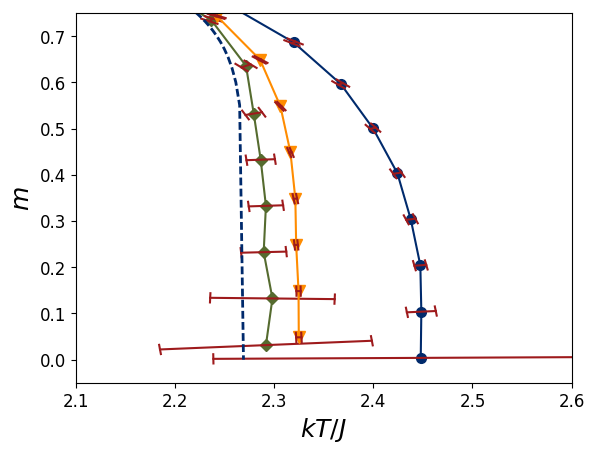}
  \caption{The critical region from Fig.~\ref{ising2}, showing results of the pseudoadaptive continuation of nontrivial steady states of the 2D Ising model using random lifting (blue circles) and first-order structure lifting (green diamonds) compared to the exact solution (dashed line). The additional line (orange triangles) represents the result of a continuation using first-order structure lifting and adaptive choice of $\tau$, $M$, and $\sigma$ with parameters $ds=0.1$, $N_{\text{fit}}=20$, $\Delta_d = 0.001$, $\theta_d = 1.0$, and spin lattice size $100\times 100$. \label{ising2.1}}
\end{figure}

\begin{figure}
  \includegraphics[width=0.9\hsize]{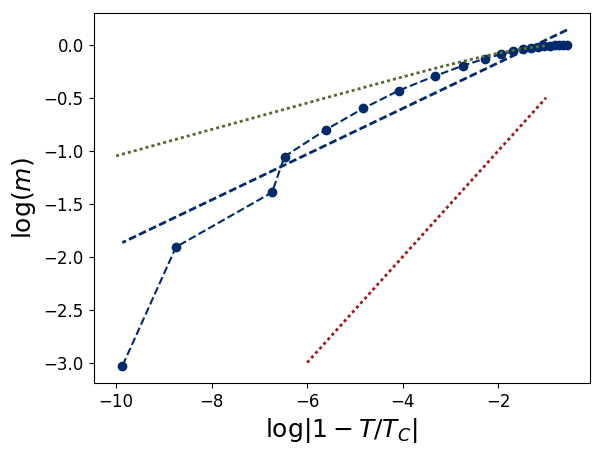}
  \caption{The data for the nontrivial $m\neq0$ branch in Fig.~\ref{ising2.1} obtained with first-order structure lifting and adaptive parameter choice (blue circles), presented in a log-log plot (natural logarithm) with a corresponding linear fit of slope $0.216$ given as blue dashed line. Note that the leftmost point is not considered for the fit. The green dotted line represents Onsager's exact solution with slope $1/8$. For comparison, the red dotted line shows a slope of $1/2$ corresponding to the mean-field solution. The required value of $T_c$ is obtained by an extrapolation of the continuation result. \label{loglogneu}}
\end{figure}

We return now to the $m\neq0$ nontrivial branch. Continuation of this using first-order structure lifting and adaptive parameter choice while using a small value of $\Delta_d$ lead to a result with small statistical errors. This result is shown in Fig.~\ref{ising2.1} using orange triangles. The curve exhibits very few fluctuations and has very small error bars. Yet, the agreement with the exact solution is not as good as the result of pseudoadaptive first-order structure lifting discussed before. Though it is still much better than the results of random lifting. Note that the error bars do not reach the exact solution. This highlights two important points. First, not surprisingly, our automatic adaptive parameter choice is not as good as the parameter choice chosen by an operator with some physical insight. Second, our error measure $\Delta$ [cf.~Eq.~(\ref{error_measure})] only represents the statistical errors of the stochastic continuation algorithm. It does not include other errors, e.g., those occurring in the lifting procedure. This implies that there are still lifting errors. Of course, our structure lifting can only be an approximation of a perfect lifting operator.

In Fig.~\ref{loglogneu} we present the results from this last continuation that has small statistical errors in a log-log plot. A linear fit then gives the critical exponent $\beta=0.216$. We see the significant deviations from the exact results close to the critical temperature and that one does not really see a clear power law. Instead, there is a range of values. However, almost all of them are much smaller than the mean-field critical exponent of $1/2$. Hence, the continuation gives a much better result than the mean-field approximation, both for $T_c$ and also for $\beta$. To obtain a clear result for the critical exponent, further improvements of the continuation technique are needed.

\section{An active Ising model\label{sec:activeIsing}}
The second example we consider is the active Ising model introduced by Solon and Tailleur in Ref.~\cite{SoTa2013prl} to capture essential features of the flocking transition occurring in many discrete and continuous models for the collective behavior of active particles \cite{ToTR2005ap, PeDB2006pre, CGGP2008epjb, Ihle2011pre, SoCT2015prl}. The active Ising model is a nonequilibrium version of a ferromagnetic model with the spin variable having the meaning of a (preferred) state of motion. Here we consider the 2D version from Ref.~\cite{SoTa2015pre} (whose notation we follow) where the particles' motion in the positive and negative $x$ direction, is influenced by the spin value. Our lattice has size $L_x\times L_y = 100\times 100$ and contains $N$ moving particles that carry a spin of $\pm 1$. Every site of the lattice $i$ can be occupied by an arbitrary number of particles, denoted by $n^{\pm}_i$ for particles with spin $\pm 1$. Consequently, for every lattice site $i$, we have a local particle density $\rho_i := n^+_i + n^-_i$ and a local magnetization $m_i := n^+_i - n^-_i$. Depending on the spin orientations, every particle can move to a neighboring lattice site or flip its spin at specific rates. The rules controlling the dynamics are as follows:
\begin{itemize}
\item A particle located at the site $i$ and having a spin $s$ flips its spin at a rate
\begin{equation}
  W(s\to-s) = \exp\left(-s\beta\frac{m_i}{\rho_i}\right),
\end{equation}
where $\beta=1/T$ is an inverse nondimensional temperature.
\item A particle with spin $s$ moves with rate $D$ to the upper or lower lattice site (unbiased motion in $y$ direction), with rate $D(1+s\varepsilon)$ to its right and with rate $D(1-s\varepsilon)$ to its left
(spin-biased motion in the $x$ direction, where $\varepsilon$ measures the strength of the bias).
\end{itemize}
With these rates, a continuous-time Markov process is defined. For the realization of the dynamics, we use a random-sequential-update algorithm. For this, we discretize the time in steps of $\Delta t$. In every step of the algorithm, a particle is chosen randomly and one of the following actions is done:
\begin{itemize}
\item With probability $W(s\to-s)\Delta t$, its spin is flipped.
\item With probability $D\Delta t$, it is moved upward.
\item With probability $D\Delta t$, it is moved downward.
\item With probability $D(1+s\varepsilon)\Delta t$, it is moved to the right.
\item With probability $D(1-s\varepsilon)\Delta t$, it is moved to the left.
\item With probability $1 - \lbrack 4D + W(s\to-s)\rbrack\Delta t$, nothing is done.
\end{itemize}
After this, the time is incremented by $\Delta t/N$. In order to keep all probabilities smaller than 1 and to minimize the probability that nothing happens, following \cite{SoTa2013prl} $\Delta t$ is chosen to be $(4D+\exp(\beta))^{-1}$.

Here, we choose $\beta=2$, $\varepsilon=0.9$ and $D=1$ and employ the developed stochastic continuation algorithm to analyze the dynamics of the system and how it depends on the average density $\rho_0=N/(L_xL_y)$, which is determined by $N$. At low densities, a trivial homogeneous unpolarized state emerges that has zero mean magnetization (sometimes called ``polarization''), i.e., $\langle m_i\rangle\approx 0$. That is, the particles show no collective behavior and the state may be seen as a homogeneous gas state. At high densities, we have $\langle m_i\rangle\neq 0$, left-right parity is broken and the particles show a collective motion either to the right or to the left. We call this the ``active liquid state.''

\begin{figure}
  \includegraphics[width=0.9\hsize]{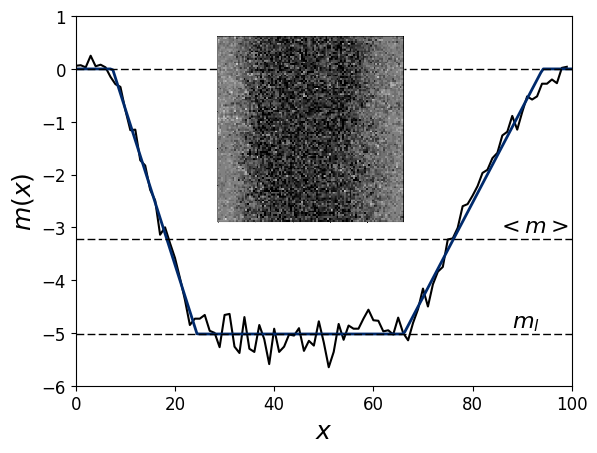}
  \caption{Example of a microscopic magnetization state of the 2D active Ising model with mean density $\rho_0=3.98$, which is in the parameter region where the gas and the liquid state coexist. The inset shows a typical snapshot of the local magnetization, clearly showing that the collective behavior results in a band structure. The central band represents the liquid phase which moves to the left through a gaseous background. The main plot gives the spatial profile of the local magnetization averaged along the band (black solid line) together with a trapezial piecewise linear fit (blue solid line). The thin dashed horizontal lines give the magnetization values of the background gas phase ($m=0$), of the band of liquid phase ($m=m_l$), and the mean $\langle m\rangle$. The fit is used to extract the macroscopic observable defined as the liquid-gas fraction $\Phi=\langle m\rangle/m_l$. The remaining parameters are $L_x=L_y=100$, $\beta=2$, $\varepsilon=0.9$, and $D=1$.
\label{GL2}}
\end{figure}

At intermediate densities, a phase separation between gas and liquid state is observed and regions of the two states coexist. In particular, one finds dense bands of moving particles as illustrated in the inset of Fig.~\ref{GL2}, which shows a snapshot of the local magnetizations. The main panel presents the spatial profile of the local magnetization averaged along the band, i.e., averaged in the $y$ direction. Employing a trapezial piecewise linear fit, we determine the mean $m_l$ of the local magnetization in the band of active liquid phase, i.e., the ``height'' of the plateau. This is then used to define the macroscopic observable
\begin{equation}
	\Phi = \frac{\langle m\rangle}{m_l} = \frac{1}{m_lL_xL_y}\sum_im_i
\end{equation}
which quantifies the liquid-gas ratio. Varying the average density $\rho_0$, the width of the band of the liquid phase changes while the value $m_l$ remains constant in the entire coexistence region.

To analyze the system behavior employing the developed stochastic continuation algorithm we define a random lifting procedure in a straightforward manner: Given the total particle number $N$, we choose random lattice sites to place the particles and assign a random spin orientation to all of them using a probability that guarantees a specific value of $\langle m\rangle$.

\begin{figure}
  \includegraphics[width=0.9\hsize]{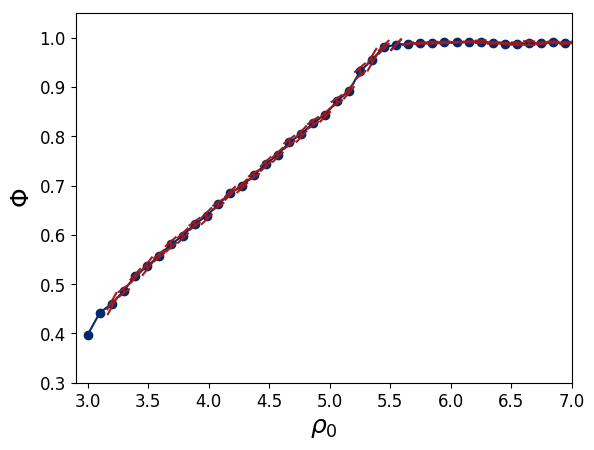}
  \caption{Bifurcation diagram of stable steady states of the 2D active Ising model obtained using stochastic continuation. The continuation is initialized on the left, uses second-order structure lifting and adaptive choices for $\tau$, $M$ and $\sigma$. The other parameters in the continuation algorithm are as follows: $ds=0.1$, $N_{\text{fit}}=15$, $\Delta_d = 0.01$, and $\theta_d = 0.7$. The (rather small) errors are indicated in red. \label{aI}}
\end{figure}

For densities belonging to coexistence states, we use a hybrid structure lifting: We modulate a coexistence state which is given by the previous continuation step by adding or removing particles at random lattice sites as well as by changing spin orientations of random particles. However, to maintain the stable coexistence structure, we only do these random adjustments in the region of the boundary between the liquid and the gas phase. The phase boundary is well defined by the sloped parts of the trapezial fits (see Fig.~\ref{GL2}). This procedure is a targeted first-order structure lifting which acts like a higher-order structure lifting. In Fig.~\ref{aI} we present continuation results for just the branches of stable states, in particular, for the transition between the banded state and the liquid state.

We initiate the continuation with a banded state obtained from a time simulation at $\rho_0=3.0$ and then employ continuation to follow the stable branch toward higher mean densities. The result in Fig.~\ref{aI} is in very good agreement with the results obtained by direct time simulation in Ref.~\cite{SoTa2015pre}. Namely the branches corresponding to the banded state and the liquid state are well approximated by straight lines and meet at a bifurcation point at $\rho_0\approx 5.4$. In particular, it seems that at comparable system sizes, the continuation approach can capture the bifurcation point much better than the direct simulations of Ref.~\cite{SoTa2015pre}. Their Fig.~10 (left) shows that for a system size of $200\times100$ there is still a jump in $\Phi$ of nearly 20\% in the region where the bifurcation point is expected. They have to go to rather large systems to well capture the bifurcation point.

Note, however, that our approach works less well when initiating the continuation at a high particle density in the liquid phase and descending in $\rho_0$. Then, when passing the bifurcation point into the coexistence region, the stable band structure needs a very long time to form. Thus, even with structure lifting, the banded stable states cannot be found with an acceptable numerical effort when passing the bifurcation point. This agrees with the general observation that the continuation method generally works better when it is initialized sufficiently far away from potential bifurcation points and then subsequently approaching them. In other words, problems should be expected when crossing a bifurcation point onto a branch having a different symmetry.

\section{Swift-Hohenberg equation\label{sec:SwiftHohenberg}}
The third and final example to illustrate the application of the developed stochastic continuation algorithm concerns a stochastic PDE, namely, the stochastic version of the Swift-Hohenberg equation in Eq.~\eqref{eq:sto_SH} \cite{Hutt2008el, HVTM1993prl}. The corresponding deterministic Swift-Hohenberg equation which is obtained when $D=0$, represents a generic model capturing the dynamics of pattern formation in the vicinity of a monotonic short-wave instability, i.e., a Turing instability \cite{CrHo1993rmp}. Depending on the control parameter $\varepsilon$, the field $\Psi$ has a variety of different stable states. For $\varepsilon<0$, the homogeneous state is stable. For a 1D system, a pitchfork bifurcation occurs at  $\varepsilon=0$ and the 
homogeneous state is unstable for $\varepsilon>0$. With the present purely cubic nonlinearity, the bifurcation is supercritical, i.e., a branch of stable periodic states emerges toward $\varepsilon>0$. Other nonlinearities may result in a subcritical bifurcation and, in general, more intricate behavior \cite{BuKn2006pre}.

\begin{figure}
  \includegraphics[width=0.9\hsize]{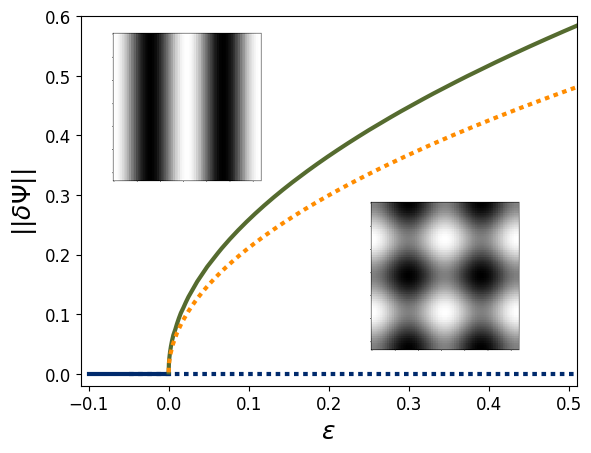}
  \caption{The bifurcation diagram for the deterministic Swift-Hohenberg equation [Eq.~\eqref{eq:sto_SH} with $D=0$] obtained by standard continuation. The solid and dashed lines are branches of stable and unstable states, respectively. The branch with $||\delta\Psi||=0$ (blue line) is the trivial, homogeneous state which changes its stability at $\varepsilon=0$, where two branches emerge supercritically.  These are a branch of stable stripe patterns (green line) and a branch of unstable square patterns (orange line). The insets show a stripe and a square state, both at $\varepsilon=0.5$.
\label{pde2path}}
\end{figure}

Here we investigate the 2D case for a square domain of size $2\pi\times 2\pi$. Then, at $\varepsilon>0$ branches of stripe patterns (equivalent to the 1D periodic states) and of square patterns simultaneously emerge supercritically. In the deterministic case one can follow them by standard continuation tools, such as \textsc{pde2path} \cite{EGUW2019springer}. Figure~\ref{pde2path} gives the corresponding bifurcation diagram employing as solution measure the $L^2$ norm of the deviations from the mean, i.e., 
\begin{equation}
||\delta\Psi|| := \sqrt{\left\langle ( \Psi - \langle\Psi\rangle )^2 \right\rangle}.
\end{equation}
Note that for the present square domain, the square patterns are always unstable while the stripes are stable. In a rectangular domain suitable for hexagonal patterns (ratio of side lengths of $\sqrt{3}$) one finds stripes and two types of hexagonal patterns and more intricate stability behavior \cite{CrHo1993rmp}.

\begin{figure}
  \includegraphics[width=0.9\hsize]{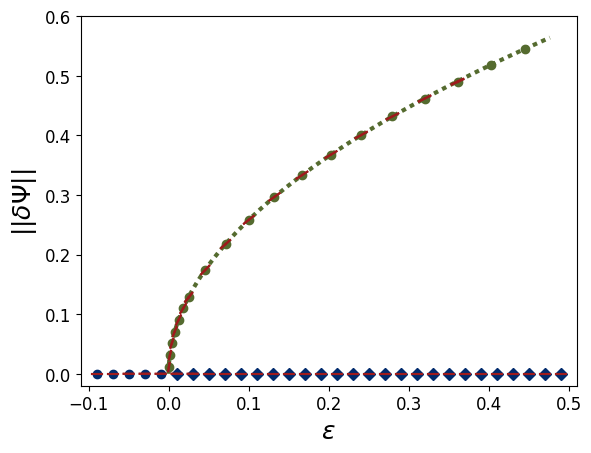}
  \caption{The bifurcation diagram obtained using stochastic continuation applied to the deterministic Swift-Hohenberg equation [Eq.~\eqref{eq:sto_SH} with $D=0$] in 2D. The circles and diamonds represent stable and unstable branches, respectively. For comparison, the dashed line gives the standard continuation results obtained with \textsc{pde2path}. For the branch of homogeneous (stripe) states, the continuation is started on the left at $\varepsilon=-0.1$ (right at $\varepsilon=0.45$). First-order structure lifting and adaptive parameter choice for $\tau$ and $\sigma$ are used. However, $M_{\text{max}}$ is fixed at 1 because no ensemble of microscopic states is needed for the deterministic system. The other parameters of the continuation algorithm are $ds=0.02$ or $ds=0.05$, $N_{\text{fit}}=6$, $\Delta_d = 0.01 ds$, and $\theta_d = 1.0$. The size of the spatially discrete representation of the field $\Psi$ is $128\times 128$.
    \label{SH1}}
\end{figure}

Although, there exist many efficient numerical continuation tools to determine bifurcation diagrams for deterministic equations such as Eq.~\eqref{eq:sto_SH} with $D=0$, here we show that stochastic continuation can also be used. This allows us to check the quality and precision of our method. In Fig.~\ref{SH1} we compare our results with the results  obtained with \textsc{pde2path}. To bring the deterministic system into a form suitable for stochastic continuation we need to define macroscopic observables and microscopic states. We take the norm $||\delta\Psi||$ as the macroscopic observable and the spatially discretized version of the field $\Psi$ as the microscopic state. We compute the microscopic dynamics via a standard pseudospectral method on a grid of $128\times 128$ points with periodic boundaries.

Here our random lifting consists in assigning a normally distributed random variable to every lattice site. By choosing a suitable variance, we obtain a random field for a specific given value of $||\delta\Psi||$. When using first-order structure lifting, we adjust the final state of the evolving procedure from the previous continuation step by multiplying it by an appropriate factor which lies near 1.

The continuation with first-order structure lifting of the branch belonging to the homogeneous state works without any problems. In this case we have a horizontal branch and the stability is directly given by the slope of the function $\widetilde{G}$. Also the branch of stripe states is obtained in good agreement with the result from standard continuation in Fig.~\ref{SH1}. However, this is only possible with the structure lifting. With random lifting, during the short-time evolution of the evolving step the Laplace operator results in diffusion on short scales and smoothes out noisy structures. Therefore, at the very beginning of the time evolution, a state generated with random lifting always moves toward a homogeneous state, allowing to follow the unstable trivial state. The slope of $\widetilde{G}$ then indicates that on short timescales the state appears to be stable.

The unstable branch belonging to the square pattern cannot automatically be followed using our structured lifting algorithm because for positive values of $\varepsilon$ it only ``sees'' the dynamics going from the unstable homogeneous solution to the stable stripe pattern. As the square pattern does not lie ``in between'' the homogeneous and stripe state, the algorithm does not detect it. However, one can devise other types of structure lifting that make it easier to find unstable solutions like the square pattern; see~Sec.~\ref{sec:conclusion_outlook} for a detailed discussion.

\begin{figure}
  \includegraphics[width=0.9\hsize]{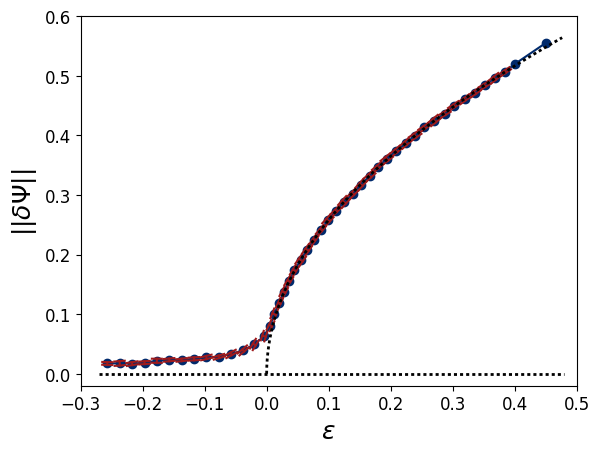}
  \caption{Results from stochastic continuation applied to the stochastic Swift-Hohenberg equation in 1D, Eq.~(\ref{eq:sh-stoch}). The continuation is initiated at $\varepsilon=0.45$, uses first-order structure lifting and fully adaptive choices for $\tau$, $M$, and $\sigma$. The other parameters of the continuation algorithm are $ds=0.02$, $N_{\text{fit}}=20$, $\Delta_d = 0.004$ and $\theta_d = 1.0$. The field $\Psi(x,t)$ is discretized on a spatial grid of $128$ cells. The dotted black line represents the continuation result for the corresponding deterministic system [Eq.~(\ref{eq:sh-stoch}) with $D=0$].
    \label{stochSH}}
\end{figure}

Next we consider the $D>0$ stochastic version of the Swift-Hohenberg equation \eqref{eq:sto_SH}. For the sake of a lower computational effort, here we restrict our attention to the 1D version of the equation with additive noise,
\begin{equation}
  \partial_t \Psi = \varepsilon\Psi - \Psi^3 - (1+\partial_x^2)^2\Psi + \sqrt{D}\eta(x,t),
  \label{eq:sh-stoch}
\end{equation}
where the noise of strength $D$ is normally distributed and uncorrelated in space and time, i.e., $\langle\eta\rangle=0$ and $\langle \eta(x',t')\eta(x,t)\rangle = \delta(x'-x)\delta(t'-t)$. We use the value $D=1$ and a domain size of $2\pi$. Again, we employ a pseudospectral method with periodic boundaries for the computation of the microscopic dynamic. Hereby, the field $\Psi(x,t)$ is discretized on a spatial grid of $128$ cells. The stochastic continuation is initiated on the stable branch of periodic states at $\varepsilon=0.45$ and proceeds toward smaller $\varepsilon$. The results are displayed in Fig.~\ref{stochSH}. We see that far from the pitchfork bifurcation of the deterministic system, the branch of periodic states is nearly not influenced by the already rather strong noise. On the other hand, the branch of stable trivial states slightly deviates from $||\delta\Psi||=0$ near the pitchfork bifurcation which has become an imperfect pitchfork bifurcation, as expected when noise is present \cite{ACLR2013pre}.

\section{Conclusion and Outlook\label{sec:conclusion_outlook}}

In the present work we have developed ways to improve the stochastic continuation of stable and unstable steady states in several aspects, mainly concerning the corrector step of the continuation algorithms. In all our work we have employed a secant predictor, however, this can be amended to any commonly used predictor. Our proposals for the corrector all concern its lifting and evolving steps. The restricting step always consists of a straight forward averaging procedure.

First, we have amended the root finding procedure for the fluctuating function $G_{\lambda}$ that in the evolving step results from the microscopic time stepper. Instead of directly applying a root finding method to the fluctuating function $G_{\lambda}$, we evaluate a set of function values in a suitable neighborhood of the initial guess provided by the predictor and perform a linear fit that is then used to determine the root. We have found that this procedure moderates the fluctuations and works in a stable way.
 
Our second improvement concerns the choice of numerical parameters. The stochastic continuation algorithm involves quite a large number of numerical parameters, e.g., the length of the microscopic time evolution in the evolving step, the number of employed microscopic realizations and the step size of the path continuation step. As their choice has a crucial influence on the quality and precision of the continuation results, we suggest to adaptively adjust specific parameters during the continuation run. In particular, in the vicinity of bifurcation points, this has led to more accurate results.

In the third amendment we have considered the lifting procedure that forms a very critical part of stochastic continuation. In most of the cases in the literature, a simple constrained random lifting is used, i.e., a microscopic state with the correct macroscopic observable imposed is chosen randomly. In contrast, we argue that one should exercise some control over its internal spatial structure that is important for the macroscopic dynamics. In particular, we have developed first- and second-order structure lifting procedures. In the case of the first-order procedure, final states from the microscopic evolution from the previous continuation step are randomly adapted to the new value of the macroscopic observable. In the case of the second-order procedure, final states from the microscopic evolution from the two previous continuation steps are used to obtain a state at the value of the macroscopic observable through random adjustments only in the region most changed over the previous continuation steps. This procedure can be further refined into an $n$th-order procedure.

Furthermore, in specific examples, it can be useful to incorporate information about particular solution behavior into the lifting and/or restricting procedure. This is done by our hybrid lifting for the active Ising model (Sec.~\ref{sec:activeIsing}) and by the weighted lifting and restriction for the agent-based model discussed in Ref.~\cite{ConsumerLockIn}.

The abilities and shortcomings of the original method and of our various proposed amendments have then be discussed using three examples: (i) the classical Ising model in 2D, (ii) an active Ising model, i.e., a kinetic Monte Carlo model for the collective behavior of active self-propelled particles, and (iii) the Swift-Hohenberg equation, a generic PDE model for short-scale pattern formation close to a Turing bifurcation. The final example has on the one hand been used to show that our stochastic continuation algorithm can be employed to continue solutions of a deterministic equation (the deterministic Swift-Hohenberg equation in 2D) and, on the other hand, to continue nontrivial steady states of a stochastic PDE (the Swift-Hohenberg equation with additive noise in 1D). These examples have shown that our improvements to the stochastic continuation method are rather important for the quality and precision of the resulting bifurcation diagrams.

We highlight that in example (i) stochastic continuation now provides a bifurcation diagram quite closely resembling Onsager's analytical solution. Our root finding procedure has allowed for a much closer approach to the bifurcation point, compare Fig.~15 of Ref.~\cite{ThLS2016pa} with the present Fig.~\ref{ising2}. Also the fluctuations in the branch of trivial states are much reduced (see Fig.~\ref{ising3}). In particular, we have obtained an estimate for the critical exponent that is much closer to the exact value than the mean-field result. However,
especially near the critical point the result is not very precise as no clear unique power law emerges. To obtain a more accurate value for the critical exponent further specific adaptations of the continuation technique are needed, e.g., one could specifically adapt the routines to look for scaling laws. In the Ising example, one could incorporate the knowledge that a scaling law exists for states close to the critical point into the method and in this way use prior information about the form of the branch. Technically, this could be done by using a scaling law with adaptive exponent for an effective determination of initial guesses for subsequent points on the branch being continued. This is, however, out of the scope of the present work.

Further improvements to the method include the lifting procedure, as the structure lifting still fails if the microscopic dynamics of the system is too slow. This we have seen in the example (ii), where it was not possible to follow the stable branch through the bifurcation point when initiating the continuation at large values of the control parameter.

Also, there can be problems when using the structure lifting to follow unstable steady states. In principle, it should be possible to follow such states, since unstable steady states of spatially extended systems normally correspond to saddle points in a very high-dimensional space of solutions. Within this space there usually exist only one or a few slow unstable modes among many fast stable modes. Therefore, on short timescales one can expect to have a healing process in the evolving step where unstable solutions are approached. This forms the basis of stochastic continuation of unstable states.

We have seen related issues in the deterministic Swift-Hohenberg system [example (iii)], where we have not been able to find the branch of the unstable square pattern, which lies between the unstable branch belonging to the homogeneous solution and the stable branch of the square pattern. The reason for this is that, during continuation with structure lifting, we let the microscopic pattern evolve on its own. As a result, the stable stripe pattern is favoured over the unstable square pattern.
To remedy this, one could once feed in the missing square pattern as a basis for the continuation of the missing branch. The idea is to use it only for the first lifting process which is followed by structure liftings as done before.

In the example of the deterministic Swift-Hohenberg equation, it is easy to obtain the missing unstable pattern from a \textsc{pde2path} solution, for example. In other examples where only stochastic continuation can be applied, the issue of how one obtains information about potentially existing unstable steady states, such as the square patterns in the Swift-Hohenberg example, is very important. An approach to this problem could be the following. As a system evolves on the way to a stable solution, the state of the system often gets close to unstable solutions, because they behave like saddle points. So, observing the time evolution, starting from an arbitrary state can give clues about possible unstable states. For example, the stripe pattern of the Swift-Hohenberg system evolves out of a homogeneous state for positive values of the parameter $\varepsilon$. But, before the stripes emerge, typically, one sees many localized peaks. These peaks are connected into stripes later on, but they are very similar to the peaks which form the square pattern.

We expect that the Eckhaus instability for a stochastic Swift-Hohenberg equation \cite{HVTM1993prl} would be another good example to learn how to deal with unstable branches as there exist many unstable branches and branches that change their stability. It would be interesting to see whether the different unstable branches can be distinguished from each other and how the method performs along a branch of stripes which changes its stability.

Here we have focused on the continuation of steady states. However, also the continuation of periodic, quasiperiodic, and chaotic dynamics may be important for the understanding of specific systems. In principle, a stochastic continuation of such solutions should be possible, if they are captured by appropriate solution measures, which clearly separate different types of solutions. Of course, in particular, for a chaotic dynamic it might be more difficult to find appropriate parameters for the continuation algorithm. Another difficulty will be the definition of a suitable lifting procedure. Molecular dynamics simulations, for example, show a chaotic dynamics and are analyzed in an equation-free framework in Ref.~\cite{FSVL2009DCDSB}. Especially, they discuss the lifting procedure and related problems.

Furthermore, our adaptive adjustments of the numerical parameters makes it possible to obtain more precise results for the zeros of the function $\widetilde{G}$. Yet, this can be improved further. Choosing the parameter values by hand in a pseudoadaptive way (as we have done for the results in Fig.~\ref{ising2}, for example) is a laborious process. However, human experience (physical insight) accumulated over time still allows one to effectively gauge the obtained plots and linear fits and to adjust parameters by hand and to obtain better results than all of the automated algorithms presented. In the future one can envisage employing machine learning, e.g., using neural networks, for this task. Currently, such methods are quite successful in the solution of visual problems.

Finally, emphasize again the connection between the slope of the function $\widetilde{G}$ and the stability of the corresponding solution branch. Moreover, we have seen that the occurrence of a bifurcation should coincide with a slope of zero. However, these connections depend on the direction of the orthogonal corrector which varies while following a curved solution branch. In the future this issue should be elaborated in detail to enable one to perform a systematic stability analysis of all states on the branches being followed.\\

The data that support the findings of this study are openly available \cite{Zenodo}.

\begin{acknowledgments}
We thank Tobias Frohoff-H\"ulsmann for helpful discussions and for providing the \textsc{pde2path} continuation runs for the deterministic Swift-Hohenberg equation.
\end{acknowledgments}

%


\begin{thebibliography}{67}%
\makeatletter
\providecommand \@ifxundefined [1]{%
 \@ifx{#1\undefined}
}%
\providecommand \@ifnum [1]{%
 \ifnum #1\expandafter \@firstoftwo
 \else \expandafter \@secondoftwo
 \fi
}%
\providecommand \@ifx [1]{%
 \ifx #1\expandafter \@firstoftwo
 \else \expandafter \@secondoftwo
 \fi
}%
\providecommand \natexlab [1]{#1}%
\providecommand \enquote  [1]{``#1''}%
\providecommand \bibnamefont  [1]{#1}%
\providecommand \bibfnamefont [1]{#1}%
\providecommand \citenamefont [1]{#1}%
\providecommand \href@noop [0]{\@secondoftwo}%
\providecommand \href [0]{\begingroup \@sanitize@url \@href}%
\providecommand \@href[1]{\@@startlink{#1}\@@href}%
\providecommand \@@href[1]{\endgroup#1\@@endlink}%
\providecommand \@sanitize@url [0]{\catcode `\\12\catcode `\$12\catcode
  `\&12\catcode `\#12\catcode `\^12\catcode `\_12\catcode `\%12\relax}%
\providecommand \@@startlink[1]{}%
\providecommand \@@endlink[0]{}%
\providecommand \url  [0]{\begingroup\@sanitize@url \@url }%
\providecommand \@url [1]{\endgroup\@href {#1}{\urlprefix }}%
\providecommand \urlprefix  [0]{URL }%
\providecommand \Eprint [0]{\href }%
\providecommand \doibase [0]{http://dx.doi.org/}%
\providecommand \selectlanguage [0]{\@gobble}%
\providecommand \bibinfo  [0]{\@secondoftwo}%
\providecommand \bibfield  [0]{\@secondoftwo}%
\providecommand \translation [1]{[#1]}%
\providecommand \BibitemOpen [0]{}%
\providecommand \bibitemStop [0]{}%
\providecommand \bibitemNoStop [0]{.\EOS\space}%
\providecommand \EOS [0]{\spacefactor3000\relax}%
\providecommand \BibitemShut  [1]{\csname bibitem#1\endcsname}%
\let\auto@bib@innerbib\@empty
\bibitem [{\citenamefont {Deutsch}\ \emph {et~al.}(2007)\citenamefont
  {Deutsch}, \citenamefont {Maini},\ and\ \citenamefont
  {Dormann}}]{DeutschMainiDormann2007}%
  \BibitemOpen
  \bibfield  {author} {\bibinfo {author} {\bibfnamefont {A.}~\bibnamefont
  {Deutsch}}, \bibinfo {author} {\bibfnamefont {P.}~\bibnamefont {Maini}}, \
  and\ \bibinfo {author} {\bibfnamefont {S.}~\bibnamefont {Dormann}},\
  }\href@noop {} {\emph {\bibinfo {title} {Cellular Automaton Modeling of
  Biological Pattern Formation: Characterization, Applications, and
  Analysis}}},\ Modeling and Simulation in Science, Engineering and Technology\
  (\bibinfo  {publisher} {Birkh{\"a}user Boston},\ \bibinfo {year}
  {2007})\BibitemShut {NoStop}%
\bibitem [{\citenamefont {Chopard}\ and\ \citenamefont
  {Droz}(2005)}]{ChopardDroz2005b}%
  \BibitemOpen
  \bibfield  {author} {\bibinfo {author} {\bibfnamefont {B.}~\bibnamefont
  {Chopard}}\ and\ \bibinfo {author} {\bibfnamefont {M.}~\bibnamefont {Droz}},\
  }\href@noop {} {\emph {\bibinfo {title} {Cellular Automata Modeling of
  Physical Systems}}},\ Al{\'e}a-Saclay\ (\bibinfo  {publisher} {Cambridge
  University Press},\ \bibinfo {address} {Cambridge},\ \bibinfo {year}
  {2005})\BibitemShut {NoStop}%
\bibitem [{\citenamefont {Tu}(1994)}]{Tu1994}%
  \BibitemOpen
  \bibfield  {author} {\bibinfo {author} {\bibfnamefont {P.}~\bibnamefont
  {Tu}},\ }\href@noop {} {\emph {\bibinfo {title} {Dynamical Systems : An
  Introduction with Applications in Economics and Biology}}}\ (\bibinfo
  {publisher} {Springer-Verlag},\ \bibinfo {address} {Berlin},\ \bibinfo {year}
  {1994})\BibitemShut {NoStop}%
\bibitem [{\citenamefont {Haken}(2006)}]{Haken2006}%
  \BibitemOpen
  \bibfield  {author} {\bibinfo {author} {\bibfnamefont {H.}~\bibnamefont
  {Haken}},\ }\href@noop {} {\emph {\bibinfo {title} {Information and
  Self-organization: A Macroscopic Approach to Complex Systems}}},\ Springer
  Series in Synergetics\ (\bibinfo  {publisher} {Springer},\ \bibinfo {address}
  {Berlin},\ \bibinfo {year} {2006})\BibitemShut {NoStop}%
\bibitem [{\citenamefont {Pismen}(2006)}]{Pismen2006}%
  \BibitemOpen
  \bibfield  {author} {\bibinfo {author} {\bibfnamefont {L.~M.}\ \bibnamefont
  {Pismen}},\ }\href@noop {} {\emph {\bibinfo {title} {Patterns and Interfaces
  in Dissipative Dynamics}}},\ Springer Series in Synergetics\ (\bibinfo
  {publisher} {Springer-Verlag},\ \bibinfo {address} {Berlin},\ \bibinfo {year}
  {2006})\BibitemShut {NoStop}%
\bibitem [{\citenamefont {Glansdorff}\ and\ \citenamefont
  {Prigogine}(1971)}]{GlansdorffPrigogine}%
  \BibitemOpen
  \bibfield  {author} {\bibinfo {author} {\bibfnamefont {P.}~\bibnamefont
  {Glansdorff}}\ and\ \bibinfo {author} {\bibfnamefont {I.}~\bibnamefont
  {Prigogine}},\ }\href@noop {} {\emph {\bibinfo {title} {Thermodynamic Theory
  of Structure, Stability and Fluctuations}}}\ (\bibinfo  {publisher} {Wiley},\
  \bibinfo {address} {New York},\ \bibinfo {year} {1971})\BibitemShut {NoStop}%
\bibitem [{\citenamefont {Kaneko}(1990)}]{PhysRevLett.65.1391}%
  \BibitemOpen
  \bibfield  {author} {\bibinfo {author} {\bibfnamefont {K.}~\bibnamefont
  {Kaneko}},\ }\href {\doibase 10.1103/PhysRevLett.65.1391} {\bibfield
  {journal} {\bibinfo  {journal} {Phys. Rev. Lett.}\ }\textbf {\bibinfo
  {volume} {65}},\ \bibinfo {pages} {1391} (\bibinfo {year}
  {1990})}\BibitemShut {NoStop}%
\bibitem [{\citenamefont {Sinha}(1992)}]{PhysRevLett.69.3306}%
  \BibitemOpen
  \bibfield  {author} {\bibinfo {author} {\bibfnamefont {S.}~\bibnamefont
  {Sinha}},\ }\href {\doibase 10.1103/PhysRevLett.69.3306} {\bibfield
  {journal} {\bibinfo  {journal} {Phys. Rev. Lett.}\ }\textbf {\bibinfo
  {volume} {69}},\ \bibinfo {pages} {3306} (\bibinfo {year}
  {1992})}\BibitemShut {NoStop}%
\bibitem [{\citenamefont {Pikovsky}\ and\ \citenamefont
  {Kurths}(1994)}]{PhysRevLett.72.1644}%
  \BibitemOpen
  \bibfield  {author} {\bibinfo {author} {\bibfnamefont {A.~S.}\ \bibnamefont
  {Pikovsky}}\ and\ \bibinfo {author} {\bibfnamefont {J.}~\bibnamefont
  {Kurths}},\ }\href {\doibase 10.1103/PhysRevLett.72.1644} {\bibfield
  {journal} {\bibinfo  {journal} {Phys. Rev. Lett.}\ }\textbf {\bibinfo
  {volume} {72}},\ \bibinfo {pages} {1644} (\bibinfo {year}
  {1994})}\BibitemShut {NoStop}%
\bibitem [{\citenamefont {Ashwin}\ \emph {et~al.}(2012)\citenamefont {Ashwin},
  \citenamefont {Wieczorek}, \citenamefont {Vitolo},\ and\ \citenamefont
  {Cox}}]{TippingPointsAWVC2012}%
  \BibitemOpen
  \bibfield  {author} {\bibinfo {author} {\bibfnamefont {P.}~\bibnamefont
  {Ashwin}}, \bibinfo {author} {\bibfnamefont {S.}~\bibnamefont {Wieczorek}},
  \bibinfo {author} {\bibfnamefont {R.}~\bibnamefont {Vitolo}}, \ and\ \bibinfo
  {author} {\bibfnamefont {P.}~\bibnamefont {Cox}},\ }\href {\doibase
  10.1098/rsta.2011.0306} {\bibfield  {journal} {\bibinfo  {journal}
  {Philosophical Transactions of the Royal Society A: Mathematical, Physical
  and Engineering Sciences}\ }\textbf {\bibinfo {volume} {370}},\ \bibinfo
  {pages} {1166} (\bibinfo {year} {2012})}\BibitemShut {NoStop}%
\bibitem [{\citenamefont {Kuznetsov}(2010)}]{Kuznetsov2010}%
  \BibitemOpen
  \bibfield  {author} {\bibinfo {author} {\bibfnamefont {Y.~A.}\ \bibnamefont
  {Kuznetsov}},\ }\href@noop {} {\emph {\bibinfo {title} {Elements of Applied
  Bifurcation Theory}}},\ \bibinfo {edition} {3rd}\ ed.\ (\bibinfo  {publisher}
  {Springer},\ \bibinfo {address} {New York},\ \bibinfo {year}
  {2010})\BibitemShut {NoStop}%
\bibitem [{\citenamefont {Strogatz}(2014)}]{Strogatz2014}%
  \BibitemOpen
  \bibfield  {author} {\bibinfo {author} {\bibfnamefont {S.~H.}\ \bibnamefont
  {Strogatz}},\ }\href@noop {} {\emph {\bibinfo {title} {Nonlinear Dynamics and
  Chaos}}}\ (\bibinfo  {publisher} {Westview Press},\ \bibinfo {address} {New
  York},\ \bibinfo {year} {2014})\BibitemShut {NoStop}%
\bibitem [{\citenamefont {Argyris}\ \emph {et~al.}(2015)\citenamefont
  {Argyris}, \citenamefont {Faust}, \citenamefont {Haase},\ and\ \citenamefont
  {Friedrich}}]{ArgyrisFaustHaaseFriedrich2015}%
  \BibitemOpen
  \bibfield  {author} {\bibinfo {author} {\bibfnamefont {J.}~\bibnamefont
  {Argyris}}, \bibinfo {author} {\bibfnamefont {G.}~\bibnamefont {Faust}},
  \bibinfo {author} {\bibfnamefont {M.}~\bibnamefont {Haase}}, \ and\ \bibinfo
  {author} {\bibfnamefont {R.}~\bibnamefont {Friedrich}},\ }\href@noop {}
  {\emph {\bibinfo {title} {An Exploration of Dynamical Systems and Chaos:
  completely revised and enlarged second edition}}}\ (\bibinfo  {publisher}
  {Springer},\ \bibinfo {address} {Berlin},\ \bibinfo {year}
  {2015})\BibitemShut {NoStop}%
\bibitem [{\citenamefont {Poincar\'{e}}(1885)}]{poincare1885}%
  \BibitemOpen
  \bibfield  {author} {\bibinfo {author} {\bibfnamefont {H.}~\bibnamefont
  {Poincar\'{e}}},\ }\href {\doibase 10.1007/BF02402204} {\bibfield  {journal}
  {\bibinfo  {journal} {Acta Math.}\ }\textbf {\bibinfo {volume} {7}},\
  \bibinfo {pages} {259} (\bibinfo {year} {1885})}\BibitemShut {NoStop}%
\bibitem [{\citenamefont {Landau}(1965)}]{LandauPhaseTransitions}%
  \BibitemOpen
  \bibfield  {author} {\bibinfo {author} {\bibfnamefont {L.}~\bibnamefont
  {Landau}},\ }in\ \href {\doibase
  https://doi.org/10.1016/B978-0-08-010586-4.50034-1} {\emph {\bibinfo
  {booktitle} {Collected Papers of L.D. Landau}}},\ \bibinfo {editor} {edited
  by\ \bibinfo {editor} {\bibfnamefont {D.}~\bibnamefont {Ter~Haar}}}\
  (\bibinfo  {publisher} {Pergamon},\ \bibinfo {address} {London},\ \bibinfo
  {year} {1965})\ pp.\ \bibinfo {pages} {193 -- 216}\BibitemShut {NoStop}%
\bibitem [{\citenamefont {Krauskopf}\ \emph {et~al.}(2007)\citenamefont
  {Krauskopf}, \citenamefont {Osinga},\ and\ \citenamefont
  {Galan-Vioque}}]{KrauskopfOsingaGalan-Vioque2007}%
  \BibitemOpen
  \bibinfo {editor} {\bibfnamefont {B.}~\bibnamefont {Krauskopf}}, \bibinfo
  {editor} {\bibfnamefont {H.~M.}\ \bibnamefont {Osinga}}, \ and\ \bibinfo
  {editor} {\bibfnamefont {J.}~\bibnamefont {Galan-Vioque}},\ eds.,\ \href@noop
  {} {\emph {\bibinfo {title} {Numerical Continuation Methods for Dynamical
  Systems}}}\ (\bibinfo  {publisher} {Springer},\ \bibinfo {address}
  {Dordrecht},\ \bibinfo {year} {2007})\BibitemShut {NoStop}%
\bibitem [{\citenamefont {Dijkstra}\ \emph {et~al.}(2014)\citenamefont
  {Dijkstra}, \citenamefont {Wubs}, \citenamefont {Cliffe}, \citenamefont
  {Doedel}, \citenamefont {Dragomirescu}, \citenamefont {Eckhardt},
  \citenamefont {Gelfgat}, \citenamefont {Hazel}, \citenamefont {Lucarini},
  \citenamefont {Salinger}, \citenamefont {Phipps}, \citenamefont
  {Sanchez-Umbria}, \citenamefont {Schuttelaars}, \citenamefont {Tuckerman},\
  and\ \citenamefont {Thiele}}]{DWCD2014ccp}%
  \BibitemOpen
  \bibfield  {author} {\bibinfo {author} {\bibfnamefont {H.~A.}\ \bibnamefont
  {Dijkstra}}, \bibinfo {author} {\bibfnamefont {F.~W.}\ \bibnamefont {Wubs}},
  \bibinfo {author} {\bibfnamefont {A.~K.}\ \bibnamefont {Cliffe}}, \bibinfo
  {author} {\bibfnamefont {E.}~\bibnamefont {Doedel}}, \bibinfo {author}
  {\bibfnamefont {I.~F.}\ \bibnamefont {Dragomirescu}}, \bibinfo {author}
  {\bibfnamefont {B.}~\bibnamefont {Eckhardt}}, \bibinfo {author}
  {\bibfnamefont {A.~Y.}\ \bibnamefont {Gelfgat}}, \bibinfo {author}
  {\bibfnamefont {A.}~\bibnamefont {Hazel}}, \bibinfo {author} {\bibfnamefont
  {V.}~\bibnamefont {Lucarini}}, \bibinfo {author} {\bibfnamefont {A.~G.}\
  \bibnamefont {Salinger}}, \bibinfo {author} {\bibfnamefont {E.~T.}\
  \bibnamefont {Phipps}}, \bibinfo {author} {\bibfnamefont {J.}~\bibnamefont
  {Sanchez-Umbria}}, \bibinfo {author} {\bibfnamefont {H.}~\bibnamefont
  {Schuttelaars}}, \bibinfo {author} {\bibfnamefont {L.~S.}\ \bibnamefont
  {Tuckerman}}, \ and\ \bibinfo {author} {\bibfnamefont {U.}~\bibnamefont
  {Thiele}},\ }\href {\doibase 10.4208/cicp.240912.180613a} {\bibfield
  {journal} {\bibinfo  {journal} {Commun. Comput. Phys.}\ }\textbf {\bibinfo
  {volume} {15}},\ \bibinfo {pages} {1} (\bibinfo {year} {2014})}\BibitemShut
  {NoStop}%
\bibitem [{\citenamefont {Allgower}\ and\ \citenamefont
  {Georg}(1987)}]{AllgowerGeorg1987}%
  \BibitemOpen
  \bibfield  {author} {\bibinfo {author} {\bibfnamefont {E.~L.}\ \bibnamefont
  {Allgower}}\ and\ \bibinfo {author} {\bibfnamefont {K.}~\bibnamefont
  {Georg}},\ }\href@noop {} {\emph {\bibinfo {title} {Introduction to Numerical
  Continuation Mmethods}}},\ Classics in Applied Mathematics\ (\bibinfo
  {publisher} {Society for Industrial Mathematics},\ \bibinfo {address}
  {Philadelphia, PA},\ \bibinfo {year} {1987})\BibitemShut {NoStop}%
\bibitem [{\citenamefont {Mei}(2000)}]{Mei2000}%
  \BibitemOpen
  \bibfield  {author} {\bibinfo {author} {\bibfnamefont {Z.}~\bibnamefont
  {Mei}},\ }\href {\doibase 10.1007/978-3-662-04177-2} {\emph {\bibinfo {title}
  {Numerical Bifurcation Analysis for Reaction-Diffusion Equations}}}\
  (\bibinfo  {publisher} {Springer},\ \bibinfo {address} {Berlin},\ \bibinfo
  {year} {2000})\BibitemShut {NoStop}%
\bibitem [{\citenamefont {Sanchez}\ \emph {et~al.}(2002)\citenamefont
  {Sanchez}, \citenamefont {Marques},\ and\ \citenamefont
  {Lopez}}]{SaML2002jcp}%
  \BibitemOpen
  \bibfield  {author} {\bibinfo {author} {\bibfnamefont {J.}~\bibnamefont
  {Sanchez}}, \bibinfo {author} {\bibfnamefont {F.}~\bibnamefont {Marques}}, \
  and\ \bibinfo {author} {\bibfnamefont {J.~M.}\ \bibnamefont {Lopez}},\ }\href
  {\doibase 10.1006/jcph.2002.7072} {\bibfield  {journal} {\bibinfo  {journal}
  {J. Comput. Phys.}\ }\textbf {\bibinfo {volume} {180}},\ \bibinfo {pages}
  {78} (\bibinfo {year} {2002})}\BibitemShut {NoStop}%
\bibitem [{\citenamefont {Sanchez~Umbria}\ and\ \citenamefont
  {Net}(2016)}]{SaNe2016epjt}%
  \BibitemOpen
  \bibfield  {author} {\bibinfo {author} {\bibfnamefont {J.}~\bibnamefont
  {Sanchez~Umbria}}\ and\ \bibinfo {author} {\bibfnamefont {M.}~\bibnamefont
  {Net}},\ }\href {\doibase 10.1140/epjst/e2015-50317-2} {\bibfield  {journal}
  {\bibinfo  {journal} {Eur. Phys. J.-Spec. Top.}\ }\textbf {\bibinfo {volume}
  {225}},\ \bibinfo {pages} {2465} (\bibinfo {year} {2016})}\BibitemShut
  {NoStop}%
\bibitem [{\citenamefont {Engelnkemper}\ \emph {et~al.}(2019)\citenamefont
  {Engelnkemper}, \citenamefont {Gurevich}, \citenamefont {Uecker},
  \citenamefont {Wetzel},\ and\ \citenamefont {Thiele}}]{EGUW2019springer}%
  \BibitemOpen
  \bibfield  {author} {\bibinfo {author} {\bibfnamefont {S.}~\bibnamefont
  {Engelnkemper}}, \bibinfo {author} {\bibfnamefont {S.~V.}\ \bibnamefont
  {Gurevich}}, \bibinfo {author} {\bibfnamefont {H.}~\bibnamefont {Uecker}},
  \bibinfo {author} {\bibfnamefont {D.}~\bibnamefont {Wetzel}}, \ and\ \bibinfo
  {author} {\bibfnamefont {U.}~\bibnamefont {Thiele}},\ }in\ \href {\doibase
  10.1007/978-3-319-91494-7_13} {\emph {\bibinfo {booktitle} {Computational
  Modeling of Bifurcations and Instabilities in Fluid Mechanics}}},\ \bibinfo
  {series and number} {Computational Methods in Applied Sciences, vol 50},\
  \bibinfo {editor} {edited by\ \bibinfo {editor} {\bibfnamefont
  {A.}~\bibnamefont {Gelfgat}}}\ (\bibinfo  {publisher} {Springer},\ \bibinfo
  {address} {Berlin},\ \bibinfo {year} {2019})\ pp.\ \bibinfo {pages}
  {459--501}\BibitemShut {NoStop}%
\bibitem [{\citenamefont {Theodoropoulos}\ \emph {et~al.}(2000)\citenamefont
  {Theodoropoulos}, \citenamefont {Qian},\ and\ \citenamefont
  {Kevrekidis}}]{ThQK2000pnasusa}%
  \BibitemOpen
  \bibfield  {author} {\bibinfo {author} {\bibfnamefont {C.}~\bibnamefont
  {Theodoropoulos}}, \bibinfo {author} {\bibfnamefont {Y.~H.}\ \bibnamefont
  {Qian}}, \ and\ \bibinfo {author} {\bibfnamefont {I.~G.}\ \bibnamefont
  {Kevrekidis}},\ }\href@noop {} {\bibfield  {journal} {\bibinfo  {journal}
  {Proc. Natl. Acad. Sci. U. S. A.}\ }\textbf {\bibinfo {volume} {97}},\
  \bibinfo {pages} {9840} (\bibinfo {year} {2000})}\BibitemShut {NoStop}%
\bibitem [{\citenamefont {Kevrekidis}\ and\ \citenamefont
  {Samaey}(2009)}]{Kev2009}%
  \BibitemOpen
  \bibfield  {author} {\bibinfo {author} {\bibfnamefont {I.~G.}\ \bibnamefont
  {Kevrekidis}}\ and\ \bibinfo {author} {\bibfnamefont {G.}~\bibnamefont
  {Samaey}},\ }\href {\doibase 10.1146/annurev.physchem.59.032607.093610}
  {\bibfield  {journal} {\bibinfo  {journal} {Ann. Rev. Phys. Chem.}\ }\textbf
  {\bibinfo {volume} {60}},\ \bibinfo {pages} {321} (\bibinfo {year}
  {2009})}\BibitemShut {NoStop}%
\bibitem [{\citenamefont {Avitabile}\ \emph {et~al.}(2014)\citenamefont
  {Avitabile}, \citenamefont {Hoyle},\ and\ \citenamefont
  {Samaey}}]{ConsumerLockIn}%
  \BibitemOpen
  \bibfield  {author} {\bibinfo {author} {\bibfnamefont {D.}~\bibnamefont
  {Avitabile}}, \bibinfo {author} {\bibfnamefont {R.~B.}\ \bibnamefont
  {Hoyle}}, \ and\ \bibinfo {author} {\bibfnamefont {G.}~\bibnamefont
  {Samaey}},\ }\href@noop {} {\bibfield  {journal} {\bibinfo  {journal} {SIAM
  J. Appl. Dyn. Syst.}\ }\textbf {\bibinfo {volume} {13}},\ \bibinfo {pages}
  {1583} (\bibinfo {year} {2014})}\BibitemShut {NoStop}%
\bibitem [{\citenamefont {Kuehn}(2012)}]{Kueh2012sjsc}%
  \BibitemOpen
  \bibfield  {author} {\bibinfo {author} {\bibfnamefont {C.}~\bibnamefont
  {Kuehn}},\ }\href {\doibase 10.1137/110839874} {\bibfield  {journal}
  {\bibinfo  {journal} {SIAM J. Sci. Comput.}\ }\textbf {\bibinfo {volume}
  {34}},\ \bibinfo {pages} {A1635} (\bibinfo {year} {2012})}\BibitemShut
  {NoStop}%
\bibitem [{\citenamefont {Thomas}\ \emph {et~al.}(2016)\citenamefont {Thomas},
  \citenamefont {Lloyd},\ and\ \citenamefont {Skeldon}}]{ThLS2016pa}%
  \BibitemOpen
  \bibfield  {author} {\bibinfo {author} {\bibfnamefont {S.~A.}\ \bibnamefont
  {Thomas}}, \bibinfo {author} {\bibfnamefont {D.~J.~B.}\ \bibnamefont
  {Lloyd}}, \ and\ \bibinfo {author} {\bibfnamefont {A.~C.}\ \bibnamefont
  {Skeldon}},\ }\href {\doibase 10.1016/j.physa.2016.07.043} {\bibfield
  {journal} {\bibinfo  {journal} {Physica A}\ }\textbf {\bibinfo {volume}
  {464}},\ \bibinfo {pages} {27} (\bibinfo {year} {2016})}\BibitemShut
  {NoStop}%
\bibitem [{\citenamefont {Barkley}\ \emph {et~al.}(2006)\citenamefont
  {Barkley}, \citenamefont {Kevrekidis},\ and\ \citenamefont
  {Stuart}}]{barkley2006}%
  \BibitemOpen
  \bibfield  {author} {\bibinfo {author} {\bibfnamefont {D.}~\bibnamefont
  {Barkley}}, \bibinfo {author} {\bibfnamefont {I.~G.}\ \bibnamefont
  {Kevrekidis}}, \ and\ \bibinfo {author} {\bibfnamefont {A.~M.}\ \bibnamefont
  {Stuart}},\ }\href {\doibase 10.1137/050638667} {\bibfield  {journal}
  {\bibinfo  {journal} {SIAM J. Appl. Dyn. Syst.}\ }\textbf {\bibinfo {volume}
  {5}},\ \bibinfo {pages} {403} (\bibinfo {year} {2006})}\BibitemShut {NoStop}%
\bibitem [{\citenamefont {Sieber}\ and\ \citenamefont
  {Krauskopf}(2007)}]{Sieber2007}%
  \BibitemOpen
  \bibfield  {author} {\bibinfo {author} {\bibfnamefont {J.}~\bibnamefont
  {Sieber}}\ and\ \bibinfo {author} {\bibfnamefont {B.}~\bibnamefont
  {Krauskopf}},\ }\href {\doibase 10.1142/S0218127407018646} {\bibfield
  {journal} {\bibinfo  {journal} {Int. J. Bifurcation Chaos}\ }\textbf
  {\bibinfo {volume} {17}},\ \bibinfo {pages} {2579} (\bibinfo {year}
  {2007})}\BibitemShut {NoStop}%
\bibitem [{\citenamefont {Sieber}\ and\ \citenamefont
  {Krauskopf}(2008)}]{Sieber2008}%
  \BibitemOpen
  \bibfield  {author} {\bibinfo {author} {\bibfnamefont {J.}~\bibnamefont
  {Sieber}}\ and\ \bibinfo {author} {\bibfnamefont {B.}~\bibnamefont
  {Krauskopf}},\ }\href {\doibase 10.1007/s11071-007-9217-2} {\bibfield
  {journal} {\bibinfo  {journal} {Nonlinear Dyn.}\ }\textbf {\bibinfo {volume}
  {51}},\ \bibinfo {pages} {365} (\bibinfo {year} {2008})}\BibitemShut
  {NoStop}%
\bibitem [{\citenamefont {Barton}\ \emph {et~al.}(2012)\citenamefont {Barton},
  \citenamefont {Mann},\ and\ \citenamefont {Burrow}}]{Barton2012}%
  \BibitemOpen
  \bibfield  {author} {\bibinfo {author} {\bibfnamefont {D.~A.}\ \bibnamefont
  {Barton}}, \bibinfo {author} {\bibfnamefont {B.~P.}\ \bibnamefont {Mann}}, \
  and\ \bibinfo {author} {\bibfnamefont {S.~G.}\ \bibnamefont {Burrow}},\
  }\href {\doibase 10.1177/1077546310384004} {\bibfield  {journal} {\bibinfo
  {journal} {J. Vib. Control}\ }\textbf {\bibinfo {volume} {18}},\ \bibinfo
  {pages} {509} (\bibinfo {year} {2012})},\ \Eprint
  {http://arxiv.org/abs/https://doi.org/10.1177/1077546310384004}
  {https://doi.org/10.1177/1077546310384004} \BibitemShut {NoStop}%
\bibitem [{\citenamefont {Frederix}\ \emph {et~al.}(2009)\citenamefont
  {Frederix}, \citenamefont {Samaey}, \citenamefont {Vandekerckhove},
  \citenamefont {Li}, \citenamefont {Nies},\ and\ \citenamefont
  {Roose}}]{FSVL2009DCDSB}%
  \BibitemOpen
  \bibfield  {author} {\bibinfo {author} {\bibfnamefont {Y.}~\bibnamefont
  {Frederix}}, \bibinfo {author} {\bibfnamefont {G.}~\bibnamefont {Samaey}},
  \bibinfo {author} {\bibfnamefont {C.}~\bibnamefont {Vandekerckhove}},
  \bibinfo {author} {\bibfnamefont {T.}~\bibnamefont {Li}}, \bibinfo {author}
  {\bibfnamefont {E.}~\bibnamefont {Nies}}, \ and\ \bibinfo {author}
  {\bibfnamefont {D.}~\bibnamefont {Roose}},\ }\href {\doibase
  10.3934/dcdsb.2009.11.855} {\bibfield  {journal} {\bibinfo  {journal}
  {Discret. Contin. Dyn. Syst.-Ser. B}\ }\textbf {\bibinfo {volume} {11}},\
  \bibinfo {pages} {855} (\bibinfo {year} {2009})}\BibitemShut {NoStop}%
\bibitem [{\citenamefont {Makeev}\ \emph {et~al.}(2002)\citenamefont {Makeev},
  \citenamefont {Maroudas},\ and\ \citenamefont {Kevrekidis}}]{MaMK2002jcp}%
  \BibitemOpen
  \bibfield  {author} {\bibinfo {author} {\bibfnamefont {A.~G.}\ \bibnamefont
  {Makeev}}, \bibinfo {author} {\bibfnamefont {D.}~\bibnamefont {Maroudas}}, \
  and\ \bibinfo {author} {\bibfnamefont {I.~G.}\ \bibnamefont {Kevrekidis}},\
  }\href {\doibase 10.1063/1.1476929} {\bibfield  {journal} {\bibinfo
  {journal} {J. Chem. Phys.}\ }\textbf {\bibinfo {volume} {116}},\ \bibinfo
  {pages} {10083} (\bibinfo {year} {2002})}\BibitemShut {NoStop}%
\bibitem [{\citenamefont {Kuehn}(2015)}]{Kueh2015sjuq}%
  \BibitemOpen
  \bibfield  {author} {\bibinfo {author} {\bibfnamefont {C.}~\bibnamefont
  {Kuehn}},\ }\href {\doibase 10.1137/140993685} {\bibfield  {journal}
  {\bibinfo  {journal} {SIAM-ASA J. Uncertain. Quantif.}\ }\textbf {\bibinfo
  {volume} {3}},\ \bibinfo {pages} {762} (\bibinfo {year} {2015})}\BibitemShut
  {NoStop}%
\bibitem [{\citenamefont {Pasupathy}\ and\ \citenamefont
  {Kim}(2011)}]{Pasupathy2011}%
  \BibitemOpen
  \bibfield  {author} {\bibinfo {author} {\bibfnamefont {R.}~\bibnamefont
  {Pasupathy}}\ and\ \bibinfo {author} {\bibfnamefont {S.}~\bibnamefont
  {Kim}},\ }\href {\doibase 10.1145/1921598.1921603} {\bibfield  {journal}
  {\bibinfo  {journal} {ACM Trans. Model. Comput. Simul.}\ }\textbf {\bibinfo
  {volume} {21}},\ \bibinfo {pages} {19} (\bibinfo {year} {2011})}\BibitemShut
  {NoStop}%
\bibitem [{\citenamefont {Sriraman}\ \emph {et~al.}(2005)\citenamefont
  {Sriraman}, \citenamefont {Kevrekidis},\ and\ \citenamefont
  {Hummer}}]{SrKH2005PRL}%
  \BibitemOpen
  \bibfield  {author} {\bibinfo {author} {\bibfnamefont {S.}~\bibnamefont
  {Sriraman}}, \bibinfo {author} {\bibfnamefont {I.~G.}\ \bibnamefont
  {Kevrekidis}}, \ and\ \bibinfo {author} {\bibfnamefont {G.}~\bibnamefont
  {Hummer}},\ }\href {\doibase 10.1103/PhysRevLett.95.130603} {\bibfield
  {journal} {\bibinfo  {journal} {Phys. Rev. Lett.}\ }\textbf {\bibinfo
  {volume} {95}},\ \bibinfo {pages} {130603} (\bibinfo {year}
  {2005})}\BibitemShut {NoStop}%
\bibitem [{\citenamefont {Kopelevich}\ \emph {et~al.}(2005)\citenamefont
  {Kopelevich}, \citenamefont {Panagiotopoulos},\ and\ \citenamefont
  {Kevrekidis}}]{KoPK2005JCP}%
  \BibitemOpen
  \bibfield  {author} {\bibinfo {author} {\bibfnamefont {D.~I.}\ \bibnamefont
  {Kopelevich}}, \bibinfo {author} {\bibfnamefont {A.~Z.}\ \bibnamefont
  {Panagiotopoulos}}, \ and\ \bibinfo {author} {\bibfnamefont {I.~G.}\
  \bibnamefont {Kevrekidis}},\ }\href {\doibase 10.1063/1.1839173} {\bibfield
  {journal} {\bibinfo  {journal} {J. Chem. Phys.}\ }\textbf {\bibinfo {volume}
  {122}},\ \bibinfo {pages} {044907} (\bibinfo {year} {2005})}\BibitemShut
  {NoStop}%
\bibitem [{\citenamefont {Sieber}\ \emph {et~al.}(2018)\citenamefont {Sieber},
  \citenamefont {Marschler},\ and\ \citenamefont {Starke}}]{Sieber2018}%
  \BibitemOpen
  \bibfield  {author} {\bibinfo {author} {\bibfnamefont {J.}~\bibnamefont
  {Sieber}}, \bibinfo {author} {\bibfnamefont {C.}~\bibnamefont {Marschler}}, \
  and\ \bibinfo {author} {\bibfnamefont {J.}~\bibnamefont {Starke}},\ }\href
  {\doibase 10.1137/17M1126084} {\bibfield  {journal} {\bibinfo  {journal}
  {SIAM J. Appl. Dyn. Syst.}\ }\textbf {\bibinfo {volume} {17}},\ \bibinfo
  {pages} {2574} (\bibinfo {year} {2018})}\BibitemShut {NoStop}%
\bibitem [{\citenamefont {Gear}\ \emph {et~al.}(2005)\citenamefont {Gear},
  \citenamefont {Kaper}, \citenamefont {Kevrekidis},\ and\ \citenamefont
  {Zagaris}}]{Kev2005}%
  \BibitemOpen
  \bibfield  {author} {\bibinfo {author} {\bibfnamefont {C.~W.}\ \bibnamefont
  {Gear}}, \bibinfo {author} {\bibfnamefont {T.~J.}\ \bibnamefont {Kaper}},
  \bibinfo {author} {\bibfnamefont {I.~G.}\ \bibnamefont {Kevrekidis}}, \ and\
  \bibinfo {author} {\bibfnamefont {A.}~\bibnamefont {Zagaris}},\ }\href
  {\doibase 10.1137/040608295} {\bibfield  {journal} {\bibinfo  {journal} {SIAM
  Journal on Applied Dynamical Systems}\ }\textbf {\bibinfo {volume} {4}},\
  \bibinfo {pages} {711} (\bibinfo {year} {2005})},\ \Eprint
  {http://arxiv.org/abs/https://doi.org/10.1137/040608295}
  {https://doi.org/10.1137/040608295} \BibitemShut {NoStop}%
\bibitem [{\citenamefont {Siettos}\ \emph {et~al.}(2003)\citenamefont
  {Siettos}, \citenamefont {Graham},\ and\ \citenamefont
  {Kevrekidis}}]{Kev2003}%
  \BibitemOpen
  \bibfield  {author} {\bibinfo {author} {\bibfnamefont {C.}~\bibnamefont
  {Siettos}}, \bibinfo {author} {\bibfnamefont {M.~D.}\ \bibnamefont {Graham}},
  \ and\ \bibinfo {author} {\bibfnamefont {I.~G.}\ \bibnamefont {Kevrekidis}},\
  }\href {\doibase 10.1063/1.1572456} {\bibfield  {journal} {\bibinfo
  {journal} {J. Chem. Phys.}\ }\textbf {\bibinfo {volume} {118}} (\bibinfo
  {year} {2003}),\ 10.1063/1.1572456}\BibitemShut {NoStop}%
\bibitem [{\citenamefont {Baxter}(2016)}]{baxter2016exactly}%
  \BibitemOpen
  \bibfield  {author} {\bibinfo {author} {\bibfnamefont {R.~J.}\ \bibnamefont
  {Baxter}},\ }\href@noop {} {\emph {\bibinfo {title} {Exactly Solved Models in
  Statistical Mechanics}}}\ (\bibinfo  {publisher} {Elsevier Science},\
  \bibinfo {address} {Amsterdam},\ \bibinfo {year} {2016})\BibitemShut
  {NoStop}%
\bibitem [{\citenamefont {Metropolis}\ \emph {et~al.}(1953)\citenamefont
  {Metropolis}, \citenamefont {Rosenbluth}, \citenamefont {Rosenbluth},
  \citenamefont {Teller},\ and\ \citenamefont {Teller}}]{MRRT1953JCP}%
  \BibitemOpen
  \bibfield  {author} {\bibinfo {author} {\bibfnamefont {N.}~\bibnamefont
  {Metropolis}}, \bibinfo {author} {\bibfnamefont {A.~W.}\ \bibnamefont
  {Rosenbluth}}, \bibinfo {author} {\bibfnamefont {M.~N.}\ \bibnamefont
  {Rosenbluth}}, \bibinfo {author} {\bibfnamefont {A.~H.}\ \bibnamefont
  {Teller}}, \ and\ \bibinfo {author} {\bibfnamefont {E.}~\bibnamefont
  {Teller}},\ }\href {\doibase 10.1063/1.1699114} {\bibfield  {journal}
  {\bibinfo  {journal} {J. Chem. Phys.}\ }\textbf {\bibinfo {volume} {21}},\
  \bibinfo {pages} {1087} (\bibinfo {year} {1953})}\BibitemShut {NoStop}%
\bibitem [{\citenamefont {Onsager}(1944)}]{Onsa1944pr}%
  \BibitemOpen
  \bibfield  {author} {\bibinfo {author} {\bibfnamefont {L.}~\bibnamefont
  {Onsager}},\ }\href {\doibase 10.1103/PhysRev.65.117} {\bibfield  {journal}
  {\bibinfo  {journal} {Phys. Rev.}\ }\textbf {\bibinfo {volume} {65}},\
  \bibinfo {pages} {117} (\bibinfo {year} {1944})}\BibitemShut {NoStop}%
\bibitem [{\citenamefont {Solon}\ and\ \citenamefont
  {Tailleur}(2015)}]{SoTa2015pre}%
  \BibitemOpen
  \bibfield  {author} {\bibinfo {author} {\bibfnamefont {A.~P.}\ \bibnamefont
  {Solon}}\ and\ \bibinfo {author} {\bibfnamefont {J.}~\bibnamefont
  {Tailleur}},\ }\href {\doibase 10.1103/PhysRevE.92.042119} {\bibfield
  {journal} {\bibinfo  {journal} {Phys. Rev. E}\ }\textbf {\bibinfo {volume}
  {92}},\ \bibinfo {pages} {042119} (\bibinfo {year} {2015})}\BibitemShut
  {NoStop}%
\bibitem [{\citenamefont {Hutt}(2008)}]{Hutt2008el}%
  \BibitemOpen
  \bibfield  {author} {\bibinfo {author} {\bibfnamefont {A.}~\bibnamefont
  {Hutt}},\ }\href {\doibase 10.1209/0295-5075/84/34003} {\bibfield  {journal}
  {\bibinfo  {journal} {Europhys. Lett.}\ }\textbf {\bibinfo {volume} {84}},\
  \bibinfo {pages} {34003} (\bibinfo {year} {2008})}\BibitemShut {NoStop}%
\bibitem [{\citenamefont {Hern{\'a}ndez-Garc{\'i}a}\ \emph
  {et~al.}(1993)\citenamefont {Hern{\'a}ndez-Garc{\'i}a}, \citenamefont
  {Vinals}, \citenamefont {Toral},\ and\ \citenamefont
  {San~Miguel}}]{HVTM1993prl}%
  \BibitemOpen
  \bibfield  {author} {\bibinfo {author} {\bibfnamefont {E.}~\bibnamefont
  {Hern{\'a}ndez-Garc{\'i}a}}, \bibinfo {author} {\bibfnamefont
  {J.}~\bibnamefont {Vinals}}, \bibinfo {author} {\bibfnamefont
  {R.}~\bibnamefont {Toral}}, \ and\ \bibinfo {author} {\bibfnamefont
  {M.}~\bibnamefont {San~Miguel}},\ }\href {\doibase
  10.1103/PhysRevLett.70.3576} {\bibfield  {journal} {\bibinfo  {journal}
  {Phys. Rev. Lett.}\ }\textbf {\bibinfo {volume} {70}},\ \bibinfo {pages}
  {3576} (\bibinfo {year} {1993})}\BibitemShut {NoStop}%
\bibitem [{\citenamefont {Vi{\~n}als}\ \emph {et~al.}(1991)\citenamefont
  {Vi{\~n}als}, \citenamefont {Hern{\'a}ndez-Garc{\'i}a}, \citenamefont
  {San~Miguel},\ and\ \citenamefont {Toral}}]{VHST1991pra}%
  \BibitemOpen
  \bibfield  {author} {\bibinfo {author} {\bibfnamefont {J.}~\bibnamefont
  {Vi{\~n}als}}, \bibinfo {author} {\bibfnamefont {E.}~\bibnamefont
  {Hern{\'a}ndez-Garc{\'i}a}}, \bibinfo {author} {\bibfnamefont
  {M.}~\bibnamefont {San~Miguel}}, \ and\ \bibinfo {author} {\bibfnamefont
  {R.}~\bibnamefont {Toral}},\ }\href {\doibase 10.1103/PhysRevA.44.1123}
  {\bibfield  {journal} {\bibinfo  {journal} {Phys. Rev. A}\ }\textbf {\bibinfo
  {volume} {44}},\ \bibinfo {pages} {1123} (\bibinfo {year}
  {1991})}\BibitemShut {NoStop}%
\bibitem [{\citenamefont {Elder}\ \emph {et~al.}(1992)\citenamefont {Elder},
  \citenamefont {Vi{\~n}als},\ and\ \citenamefont {Grant}}]{ElVG1992prl}%
  \BibitemOpen
  \bibfield  {author} {\bibinfo {author} {\bibfnamefont {K.~R.}\ \bibnamefont
  {Elder}}, \bibinfo {author} {\bibfnamefont {J.}~\bibnamefont {Vi{\~n}als}}, \
  and\ \bibinfo {author} {\bibfnamefont {M.}~\bibnamefont {Grant}},\
  }\href@noop {} {\bibfield  {journal} {\bibinfo  {journal} {Phys. Rev. Lett.}\
  }\textbf {\bibinfo {volume} {68}},\ \bibinfo {pages} {3024} (\bibinfo {year}
  {1992})}\BibitemShut {NoStop}%
\bibitem [{\citenamefont {Agez}\ \emph {et~al.}(2013)\citenamefont {Agez},
  \citenamefont {Clerc}, \citenamefont {Louvergneaux},\ and\ \citenamefont
  {Rojas}}]{ACLR2013pre}%
  \BibitemOpen
  \bibfield  {author} {\bibinfo {author} {\bibfnamefont {G.}~\bibnamefont
  {Agez}}, \bibinfo {author} {\bibfnamefont {M.~G.}\ \bibnamefont {Clerc}},
  \bibinfo {author} {\bibfnamefont {E.}~\bibnamefont {Louvergneaux}}, \ and\
  \bibinfo {author} {\bibfnamefont {R.~G.}\ \bibnamefont {Rojas}},\ }\href
  {\doibase 10.1103/PhysRevE.87.042919} {\bibfield  {journal} {\bibinfo
  {journal} {Phys. Rev. E}\ }\textbf {\bibinfo {volume} {87}},\ \bibinfo
  {pages} {042919} (\bibinfo {year} {2013})}\BibitemShut {NoStop}%
\bibitem [{\citenamefont {Taneike}\ \emph {et~al.}(2002)\citenamefont
  {Taneike}, \citenamefont {Nihei},\ and\ \citenamefont {Shiwa}}]{TaNS2002pla}%
  \BibitemOpen
  \bibfield  {author} {\bibinfo {author} {\bibfnamefont {T.}~\bibnamefont
  {Taneike}}, \bibinfo {author} {\bibfnamefont {T.}~\bibnamefont {Nihei}}, \
  and\ \bibinfo {author} {\bibfnamefont {Y.}~\bibnamefont {Shiwa}},\ }\href
  {\doibase 10.1016/S0375-9601(02)01257-4} {\bibfield  {journal} {\bibinfo
  {journal} {Phys. Lett. A}\ }\textbf {\bibinfo {volume} {303}},\ \bibinfo
  {pages} {212} (\bibinfo {year} {2002})}\BibitemShut {NoStop}%
\bibitem [{\citenamefont {Doedel}\ \emph
  {et~al.}(1991{\natexlab{a}})\citenamefont {Doedel}, \citenamefont {Keller},\
  and\ \citenamefont {Kernevez}}]{Doedel1}%
  \BibitemOpen
  \bibfield  {author} {\bibinfo {author} {\bibfnamefont {E.}~\bibnamefont
  {Doedel}}, \bibinfo {author} {\bibfnamefont {H.~B.}\ \bibnamefont {Keller}},
  \ and\ \bibinfo {author} {\bibfnamefont {J.~P.}\ \bibnamefont {Kernevez}},\
  }\href {\doibase 10.1142/S0218127491000397} {\bibfield  {journal} {\bibinfo
  {journal} {Int. J. Bifurcation Chaos}\ }\textbf {\bibinfo {volume} {1}},\
  \bibinfo {pages} {493} (\bibinfo {year} {1991}{\natexlab{a}})}\BibitemShut
  {NoStop}%
\bibitem [{AUT(2012)}]{AUTO}%
  \BibitemOpen
  \href@noop {} {\emph {\bibinfo {title} {Auto-07p: Continuation and
  Bifurcation Software for Ordinary Differential Equations}}}\ (\bibinfo
  {publisher} {Concordia University},\ \bibinfo {address} {Montreal, Canada},\
  \bibinfo {year} {2012})\BibitemShut {NoStop}%
\bibitem [{\citenamefont {Doedel}\ \emph
  {et~al.}(1991{\natexlab{b}})\citenamefont {Doedel}, \citenamefont {Keller},\
  and\ \citenamefont {Kernevez}}]{Doedel2}%
  \BibitemOpen
  \bibfield  {author} {\bibinfo {author} {\bibfnamefont {E.}~\bibnamefont
  {Doedel}}, \bibinfo {author} {\bibfnamefont {H.~B.}\ \bibnamefont {Keller}},
  \ and\ \bibinfo {author} {\bibfnamefont {J.~P.}\ \bibnamefont {Kernevez}},\
  }\href {\doibase 10.1142/S0218127491000555} {\bibfield  {journal} {\bibinfo
  {journal} {Int. J. Bifurcation Chaos}\ }\textbf {\bibinfo {volume} {1}},\
  \bibinfo {pages} {745} (\bibinfo {year} {1991}{\natexlab{b}})}\BibitemShut
  {NoStop}%
\bibitem [{\citenamefont {Uecker}\ \emph {et~al.}(2014)\citenamefont {Uecker},
  \citenamefont {Wetzel},\ and\ \citenamefont {Rademacher}}]{pde2path}%
  \BibitemOpen
  \bibfield  {author} {\bibinfo {author} {\bibfnamefont {H.}~\bibnamefont
  {Uecker}}, \bibinfo {author} {\bibfnamefont {D.}~\bibnamefont {Wetzel}}, \
  and\ \bibinfo {author} {\bibfnamefont {J.~D.~M.}\ \bibnamefont
  {Rademacher}},\ }\href {\doibase 10.4208/nmtma.2014.1231nm} {\bibfield
  {journal} {\bibinfo  {journal} {Numer. Math. Theor. Meth. Appl.}\ }\textbf
  {\bibinfo {volume} {7}},\ \bibinfo {pages} {58} (\bibinfo {year}
  {2014})}\BibitemShut {NoStop}%
\bibitem [{\citenamefont {Yang}(1952)}]{Yang1952}%
  \BibitemOpen
  \bibfield  {author} {\bibinfo {author} {\bibfnamefont {C.~N.}\ \bibnamefont
  {Yang}},\ }\href {\doibase 10.1103/PhysRev.85.808} {\bibfield  {journal}
  {\bibinfo  {journal} {Phys. Rev.}\ }\textbf {\bibinfo {volume} {85}},\
  \bibinfo {pages} {808} (\bibinfo {year} {1952})}\BibitemShut {NoStop}%
\bibitem [{\citenamefont {Reichl}(2016)}]{reichl1999modern}%
  \BibitemOpen
  \bibfield  {author} {\bibinfo {author} {\bibfnamefont {L.~E.}\ \bibnamefont
  {Reichl}},\ }\href@noop {} {\emph {\bibinfo {title} {A Modern Course in
  Statistical Physics}}}\ (\bibinfo  {publisher} {Wiley-VCH},\ \bibinfo
  {address} {Weinheim},\ \bibinfo {year} {2016})\BibitemShut {NoStop}%
\bibitem [{\citenamefont {Hughes}\ \emph {et~al.}(2014)\citenamefont {Hughes},
  \citenamefont {Thiele},\ and\ \citenamefont
  {Archer}}]{hughes2014introduction}%
  \BibitemOpen
  \bibfield  {author} {\bibinfo {author} {\bibfnamefont {A.~P.}\ \bibnamefont
  {Hughes}}, \bibinfo {author} {\bibfnamefont {U.}~\bibnamefont {Thiele}}, \
  and\ \bibinfo {author} {\bibfnamefont {A.~J.}\ \bibnamefont {Archer}},\
  }\href@noop {} {\bibfield  {journal} {\bibinfo  {journal} {Am. J. Phys.}\
  }\textbf {\bibinfo {volume} {82}},\ \bibinfo {pages} {1119} (\bibinfo {year}
  {2014})}\BibitemShut {NoStop}%
\bibitem [{\citenamefont {Ferdinand}\ and\ \citenamefont
  {Fisher}(1969)}]{IsingFiniteSize}%
  \BibitemOpen
  \bibfield  {author} {\bibinfo {author} {\bibfnamefont {A.~E.}\ \bibnamefont
  {Ferdinand}}\ and\ \bibinfo {author} {\bibfnamefont {M.~E.}\ \bibnamefont
  {Fisher}},\ }\href {\doibase 10.1103/PhysRev.185.832} {\bibfield  {journal}
  {\bibinfo  {journal} {Phys. Rev.}\ }\textbf {\bibinfo {volume} {185}},\
  \bibinfo {pages} {832} (\bibinfo {year} {1969})}\BibitemShut {NoStop}%
\bibitem [{\citenamefont {Solon}\ and\ \citenamefont
  {Tailleur}(2013)}]{SoTa2013prl}%
  \BibitemOpen
  \bibfield  {author} {\bibinfo {author} {\bibfnamefont {A.~P.}\ \bibnamefont
  {Solon}}\ and\ \bibinfo {author} {\bibfnamefont {J.}~\bibnamefont
  {Tailleur}},\ }\href {\doibase 10.1103/physrevlett.111.078101} {\bibfield
  {journal} {\bibinfo  {journal} {Phys. Rev. Lett.}\ }\textbf {\bibinfo
  {volume} {111}},\ \bibinfo {pages} {078101} (\bibinfo {year}
  {2013})}\BibitemShut {NoStop}%
\bibitem [{\citenamefont {Toner}\ \emph {et~al.}(2005)\citenamefont {Toner},
  \citenamefont {Tu},\ and\ \citenamefont {Ramaswamy}}]{ToTR2005ap}%
  \BibitemOpen
  \bibfield  {author} {\bibinfo {author} {\bibfnamefont {J.}~\bibnamefont
  {Toner}}, \bibinfo {author} {\bibfnamefont {Y.~H.}\ \bibnamefont {Tu}}, \
  and\ \bibinfo {author} {\bibfnamefont {S.}~\bibnamefont {Ramaswamy}},\ }\href
  {\doibase 10.1016/j.aop.2005.04.011} {\bibfield  {journal} {\bibinfo
  {journal} {Ann. Phys.}\ }\textbf {\bibinfo {volume} {318}},\ \bibinfo {pages}
  {170} (\bibinfo {year} {2005})}\BibitemShut {NoStop}%
\bibitem [{\citenamefont {Peruani}\ \emph {et~al.}(2006)\citenamefont
  {Peruani}, \citenamefont {Deutsch},\ and\ \citenamefont
  {B{\"a}r}}]{PeDB2006pre}%
  \BibitemOpen
  \bibfield  {author} {\bibinfo {author} {\bibfnamefont {F.}~\bibnamefont
  {Peruani}}, \bibinfo {author} {\bibfnamefont {A.}~\bibnamefont {Deutsch}}, \
  and\ \bibinfo {author} {\bibfnamefont {M.}~\bibnamefont {B{\"a}r}},\ }\href
  {\doibase 10.1103/PhysRevE.74.030904} {\bibfield  {journal} {\bibinfo
  {journal} {Phys. Rev. E}\ }\textbf {\bibinfo {volume} {74}},\ \bibinfo
  {pages} {030904(R)} (\bibinfo {year} {2006})}\BibitemShut {NoStop}%
\bibitem [{\citenamefont {Chat{\'e}}\ \emph {et~al.}(2008)\citenamefont
  {Chat{\'e}}, \citenamefont {Ginelli}, \citenamefont {Gr{\'e}goire},
  \citenamefont {Peruani},\ and\ \citenamefont {Raynaud}}]{CGGP2008epjb}%
  \BibitemOpen
  \bibfield  {author} {\bibinfo {author} {\bibfnamefont {H.}~\bibnamefont
  {Chat{\'e}}}, \bibinfo {author} {\bibfnamefont {F.}~\bibnamefont {Ginelli}},
  \bibinfo {author} {\bibfnamefont {G.}~\bibnamefont {Gr{\'e}goire}}, \bibinfo
  {author} {\bibfnamefont {F.}~\bibnamefont {Peruani}}, \ and\ \bibinfo
  {author} {\bibfnamefont {F.}~\bibnamefont {Raynaud}},\ }\href {\doibase
  10.1140/epjb/e2008-00275-9} {\bibfield  {journal} {\bibinfo  {journal} {Eur.
  Phys. J. B}\ }\textbf {\bibinfo {volume} {64}},\ \bibinfo {pages} {451}
  (\bibinfo {year} {2008})}\BibitemShut {NoStop}%
\bibitem [{\citenamefont {Ihle}(2011)}]{Ihle2011pre}%
  \BibitemOpen
  \bibfield  {author} {\bibinfo {author} {\bibfnamefont {T.}~\bibnamefont
  {Ihle}},\ }\href {\doibase 10.1103/PhysRevE.83.030901} {\bibfield  {journal}
  {\bibinfo  {journal} {Phys. Rev. E}\ }\textbf {\bibinfo {volume} {83}},\
  \bibinfo {pages} {030901(R)} (\bibinfo {year} {2011})}\BibitemShut {NoStop}%
\bibitem [{\citenamefont {Solon}\ \emph {et~al.}(2015)\citenamefont {Solon},
  \citenamefont {Chat{\'e}},\ and\ \citenamefont {Tailleur}}]{SoCT2015prl}%
  \BibitemOpen
  \bibfield  {author} {\bibinfo {author} {\bibfnamefont {A.~P.}\ \bibnamefont
  {Solon}}, \bibinfo {author} {\bibfnamefont {H.}~\bibnamefont {Chat{\'e}}}, \
  and\ \bibinfo {author} {\bibfnamefont {J.}~\bibnamefont {Tailleur}},\ }\href
  {\doibase 10.1103/PhysRevLett.114.068101} {\bibfield  {journal} {\bibinfo
  {journal} {Phys. Rev. Lett.}\ }\textbf {\bibinfo {volume} {114}},\ \bibinfo
  {pages} {068101} (\bibinfo {year} {2015})}\BibitemShut {NoStop}%
\bibitem [{\citenamefont {Cross}\ and\ \citenamefont
  {Hohenberg}(1993)}]{CrHo1993rmp}%
  \BibitemOpen
  \bibfield  {author} {\bibinfo {author} {\bibfnamefont {M.~C.}\ \bibnamefont
  {Cross}}\ and\ \bibinfo {author} {\bibfnamefont {P.~C.}\ \bibnamefont
  {Hohenberg}},\ }\href {\doibase 10.1103/RevModPhys.65.851} {\bibfield
  {journal} {\bibinfo  {journal} {Rev. Mod. Phys.}\ }\textbf {\bibinfo {volume}
  {65}},\ \bibinfo {pages} {851} (\bibinfo {year} {1993})}\BibitemShut
  {NoStop}%
\bibitem [{\citenamefont {Burke}\ and\ \citenamefont
  {Knobloch}(2006)}]{BuKn2006pre}%
  \BibitemOpen
  \bibfield  {author} {\bibinfo {author} {\bibfnamefont {J.}~\bibnamefont
  {Burke}}\ and\ \bibinfo {author} {\bibfnamefont {E.}~\bibnamefont
  {Knobloch}},\ }\href {\doibase 10.1103/PhysRevE.73.056211} {\bibfield
  {journal} {\bibinfo  {journal} {Phys. Rev. E}\ }\textbf {\bibinfo {volume}
  {73}},\ \bibinfo {pages} {056211} (\bibinfo {year} {2006})}\BibitemShut
  {NoStop}%
\bibitem [{\citenamefont {Willers}\ \emph {et~al.}(2020)\citenamefont
  {Willers}, \citenamefont {Thiele}, \citenamefont {Archer}, \citenamefont
  {Lloyd},\ and\ \citenamefont {Kamps}}]{Zenodo}%
  \BibitemOpen
  \bibfield  {author} {\bibinfo {author} {\bibfnamefont {C.}~\bibnamefont
  {Willers}}, \bibinfo {author} {\bibfnamefont {U.}~\bibnamefont {Thiele}},
  \bibinfo {author} {\bibfnamefont {A.~J.}\ \bibnamefont {Archer}}, \bibinfo
  {author} {\bibfnamefont {D.~J.~B.}\ \bibnamefont {Lloyd}}, \ and\ \bibinfo
  {author} {\bibfnamefont {O.}~\bibnamefont {Kamps}},\ }\href@noop {} {\emph
  {\bibinfo {title} {Data Supplement for ``Adaptive stochastic continuation
  with a modified lifting procedure applied to complex systems''}}},\
  https://doi.org/10.5281/zenodo.4002055\ (\bibinfo {year} {2020})\BibitemShut
  {NoStop}%
\end{thebibliography}
\end{document}